\documentclass[journal]{IEEEtran}

\ifCLASSINFOpdf
  \usepackage[pdftex]{graphicx}
\else
\fi

\usepackage{cite}
\usepackage[pdftex]{graphicx}
\DeclareGraphicsExtensions{.pdf,.jpeg,.png}

\usepackage{amsmath}
\usepackage{amssymb}
\usepackage{mathtools}
\usepackage{siunitx}

\usepackage{xcolor}
\usepackage[normalem]{ulem}
\usepackage{multirow}

\usepackage{verbatim}
\usepackage{soul}
\usepackage{pifont}
\newcommand{\cmark}{\ding{51}}%
\newcommand{\xmark}{\ding{55}}%
\begin{document}

\title{\color{black}Process Variation-Aware Compact Model of Strip Waveguides for Photonic Circuit Simulation}

\author{Aneek~James,
        Anthony~Rizzo,
        Yuyang~Wang,
        Asher~Novick,
        Songli~Wang,
        Robert~Parsons,
        Kaylx~Jang,
        Maarten~Hattink,
        and~Keren~Bergman

\thanks{This work was supported in part by the U.S. Advanced Research Projects Agency--Energy under ENLITENED Grant DE-AR000843 and in part by the U.S. Defense Advanced Research Projects Agency under PIPES Grant HR00111920014.} 

\thanks{A. James, Y. Wang, A. Novick, S. Wang, R. Parsons, K. Jang, M. Hattink, and K. Bergman are with the Department of Electrical Engineering, Columbia University, New York, NY 10027, USA. (Corresponding author: Aneek James, e-mail: aej2149@columbia.edu).}
\thanks{A. Rizzo is with the Air Force Research Laboratory Information Directorate, Rome, NY 13441, USA.}
\thanks{© 2023 IEEE. Personal use of this material is permitted. Permission from IEEE must be obtained for all other uses, in any current or future media, including reprinting/republishing this material for advertising or promotional purposes, creating new collective works, for resale or redistribution to servers or lists, or reuse of any copyrighted component of this work in other works.}
}
\maketitle

\begin{abstract}
\color{black}We report a novel process variation-aware compact model of strip waveguides that is suitable for circuit-level simulation of waveguide-based process design kit (PDK) elements. The model is shown to describe both loss and---using a novel expression for the thermo-optic effect in high index contrast materials---the thermo-optic behavior of strip waveguides. A novel group extraction method enables modeling the effective index's ($n_{\mathrm{eff}}$) sensitivity to local process variations without the presumption of variation source. Use of Euler-bend Mach-Zehnder interferometers (MZIs) fabricated in a 300~mm wafer run allow model parameter extraction at widths up to 2.5~$\mu$m  (highly multi-mode) with strong suppression of higher-order mode excitation. Experimental results prove the reported model can self-consistently describe waveguide phase, loss, and thermo-optic behavior across all measured devices over an unprecedented range of optical bandwidth, waveguide widths, and temperatures.\color{black}

\end{abstract}


\IEEEpeerreviewmaketitle
\section{Introduction}
\IEEEPARstart{S}{ilicon} photonics \color{black}(SiPh) \color{black} has seen explosive growth in demand as a technology platform, driven by its adoption in data centers (DC), high performance computing (HPC) \cite{rizzo2022petabit, Cheng:18, wade2020teraphy}, quantum computing \cite{moody20222022, steinbrecher2019quantum, takeda2019toward, harris2016large, bourassa2021blueprint}, and radio-frequency communication systems \cite{marpaung2019integrated, yang2020terahertz, zong20196g}. \color{black}SiPh's \color{black} rapid rise and maturation has been enabled by its ability to leverage decades of research in the complementary metal–oxide–semiconductor (CMOS) industry, drastically reducing the typical research and development (R\&D) costs associated with new semiconductor technologies \cite{shi2020scaling,zhou2018development,sabella2019silicon}. \color{black}SiPh\color{black}, however, has not yet been able to mimic CMOS yield prediction tools for evaluating photonic integrated circuits (PICs). Yield is a ubiquitous metric used across semiconductor manufacturing, with improvements in yield being strongly correlated with reductions in the time and costs associated with PIC design cycles \cite{gardner2000solving, kumar2006review, bogaerts2018silicon}.  \color{black}The need for predictive yield models can be mitigated to some degree \color{black} by designing variation-robust devices \cite{Rizzo:23} or PICs such that performance variations can be tolerated or corrected for post fabrication \cite{Wang:21, krishnamoorthy2011exploiting}. \color{black}In each of these cases, however, quantitative yield data cannot be determined  prior to fabrication---an obstacle that will be exacerbated as the number of components per PIC in silicon is projected to scale well into the millions within the next decade \cite{margalit2021perspective}. \color{black} Circuit designers also need tools to optimize system-level performance through device-level design choices \cite{wang2023dispersion}. To meet rising circuit design complexity, commercial foundries must develop process design kits (PDKs) that include compact models that are both parameterized over a wide range of relevant design and environmental variables and describe all important device figures of merit \cite{woltjer2006industrial, moezi2012statistical}. It is essential that strip waveguides in particular---a critical component of most SiPh circuits---are accurately modeled according to their expected fabricated performance. \color{black}
\begin{table}
\color{black}\caption{Features for modeling strip waveguide performance. The model in this work describes phase, loss and thermal behavior effects over a broad range of wavelengths and waveguide geometries.}
    \centering
    \color{black}\begin{tabular}{cccc}
        \hline
       Model Features & \cite{xing2018accurate} & \cite{lu2017performance} & \textbf{This Work} \\ \hline
        Wavelength [nm] & 1550 & 1520--1570 & \textbf{1450--1650}\\
        Nominal Width Range [nm] & 480 & 480--500 & \textbf{400--2500}\\
        Considered Variation Sources & $w$,$t$ & $w$,$t$ & \textbf{Arbitrary}\\
        Statistical Parameter Variations & \cmark & \cmark & \cmark\\
        Waveguide Scattering Losses & \xmark & \xmark & \cmark\\
        Thermo-optic Effect & \xmark & \xmark & \cmark\\\hline
        \multicolumn{4}{@{}l}{$w$ - Waveguide Width Variations}\\
        \multicolumn{4}{@{}l}{$t$ - Waveguide Thickness Variations}
    \end{tabular}
    \label{tab:state_of_art}
\end{table}

\color{black}
Broadly speaking, there are three ways to construct compact models: (i) look up table-based models, obtained directly from measurements or device simulations, (ii) models based on empirical fit functions, and (iii) physics-based models \cite{woltjer2006industrial}. Most previously reported work falls under the look-up table-based category \cite{zortman2010silicon,lu2017performance, xing2018accurate, zhang2022inference, stievater2022optical}. These models can be parameterized using look-up tables (LUTs), where interpolation is used to predict the performance of designs not explicitly defined in the table. Ensuring that LUT models are accurate over a wide range of input parameters, however, requires measuring all waveguide figures of merit for every combination of input parameters; a task that scales exponentially with the number of modeled independent variables. Prior demonstrations methods also require the explicit connection of the measured effective and group index variations to a predefined number of process variation sources, introducing the possibility of error if any systemic deviations exist between the simulation configuration and the realities of the fabrication process.

\color{black}

In this paper, we report to the best of our knowledge, the first geometry-parameterized compact model of strip waveguides that can capture device performance over a wide range of wavelengths and waveguide geometries (see Table \ref{tab:state_of_art}). Using a novel derivation of the thermo-optic effect that is accurate for high-index contrast waveguides, we demonstrate our model's ability to describe both scattering loss and the thermo-optic effect as a function of both design and statistical parameters. A novel group-extraction-based method allows the characterization of process variations without presumption of a source or its associated sensitivity. This extraction methodology is used to construct a model from dozens of geometric variations of Mach-Zehnder Interferometers (MZIs) fabricated in a 300~mm commercial foundry. These use of Euler bends in these MZIs permits the characterization of wide waveguide performance with minimal higher-order mode excitation. Experimental results validate the model's accuracy in describing the phase, loss, and thermo-optic performance across the entire wafer. The model is also implemented in Verilog-A to demonstrate compatibility with electronic-photonic co-simulation environments \cite{shawon2020rapid, zhang2017compact, sorace2015electro}. This work represents a key step toward the modeling of waveguide-based PDK components, enabling true-to-measurement circuit simulation at massive integration densities.
\color{black}

\color{black}\section{Physics-Aware Model Development}\label{sec:model}\color{black}
\begin{figure}
    \centering
    \includegraphics{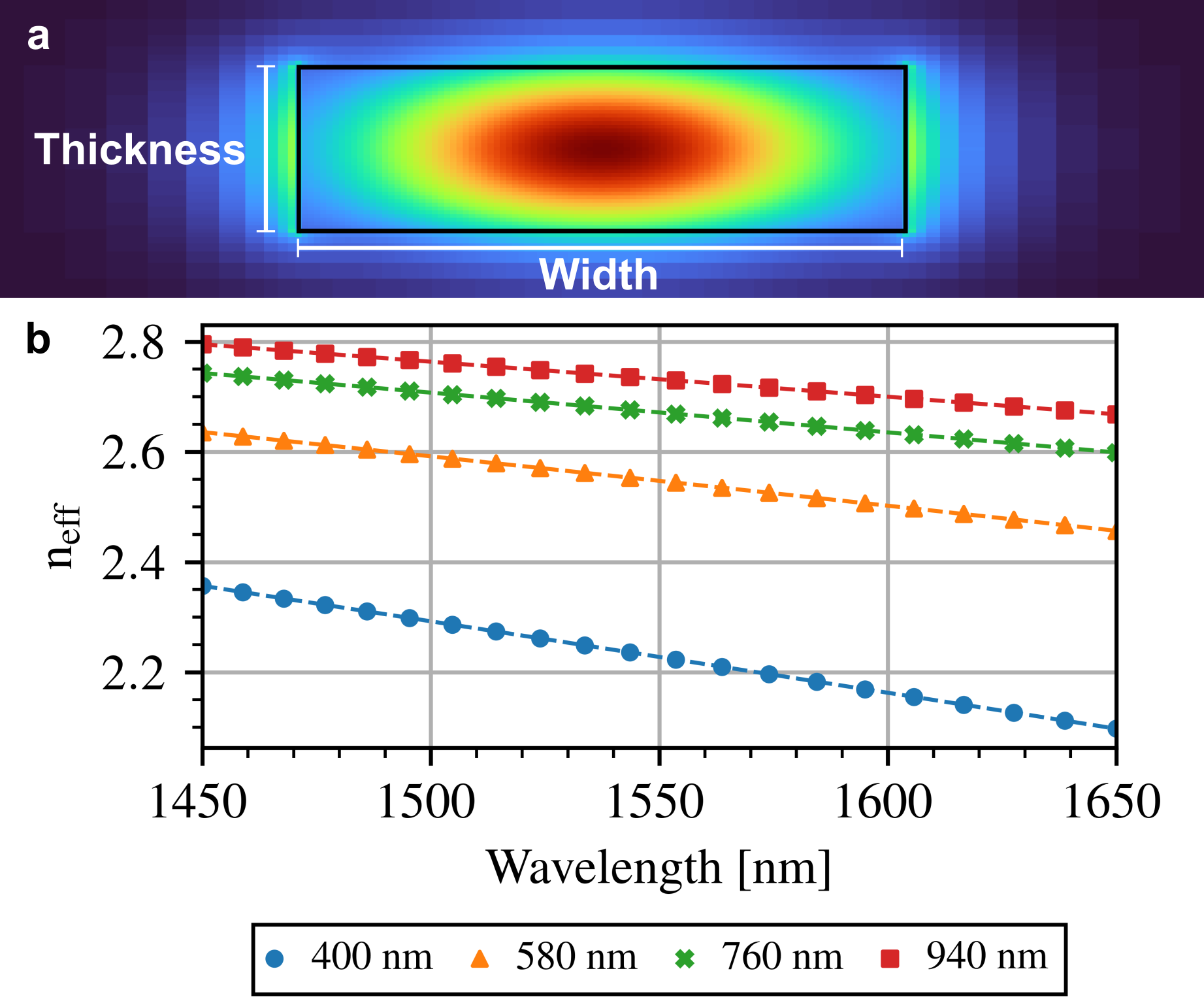}
    \caption{\textbf{a,} Example electric field profile taken from Lumerical MODE. \textbf{b,} Simulated (scatter) and modeled (dashed) effective index vs wavelength for several waveguide widths. Each waveguide was simulated with a thickness of 220~nm.}
    \label{fig:fit_taylor_validation}
\end{figure}
Because the mode condition of an optical waveguide is described via a transcendental equation, \color{black}completely generalized \color{black} analytical solutions for the effective index \color{black}$\left(n_{\text{eff}}\right)$ \color{black} are impossible to derive \cite{saleh2019fundamentals}. We therefore propose, as discussed in \cite{james2022flexible}, finding a behavioral model that accurately captures its dependence on all design parameters over \color{black}the relevant \color{black} ranges of interest. \color{black}In this section, we develop dependency models for the design parameters available. The semi-physical nature of the model is then leveraged to describe both the scattering loss and the thermo-optic coefficient. Process variations, whether of a design parameter or not, will be covered in Section \ref{sec:process_aware}.\color{black}
\\
\subsection{Wavelength Dependence}\color{black}
The wavelength dependence of the waveguide \color{black}$n_{\text{eff}}$ \color{black} is first considered. The \color{black}$n_{\text{eff}}$ \color{black} of several silicon-on-insulator (SOI) waveguide geometries were simulated in Lumerical MODE (Fig. \ref{fig:fit_taylor_validation}a). From the results, it is shown that the wavelength dependence over the S-, C-, and L-bands for all geometries is well-approximated by a second-order Taylor expansion for a wide range of waveguide widths sufficiently above the cutoff condition (Fig. \ref{fig:fit_taylor_validation}b): 
\begin{equation}\label{eq:effective_index_model}
    \color{black}n_{\text{eff, model}}(\lambda)=\sum_{i=0}^2\frac{1}{i!}\left.\frac{\partial^in_{\text{eff}}}{\partial\lambda^i}\right|_{\lambda=\lambda_0}(\lambda-\lambda_0)^i.
\end{equation}
\color{black}\subsection{Geometric Dependence}\color{black}
As the Taylor expansion only captures the wavelength-dependence, it is clear that the fitting parameters $\partial^2n_{\text{eff}}/\partial\lambda^2$, $\partial n_{\text{eff}}/\partial\lambda$ and $\partial^0n_{\text{eff}}/\partial\lambda^0$ (hereafter referred to as $n_{\text{\text{eff}}, 0}$) are responsible for capturing the dependence on waveguide geometry. With respect to width, all three fitting parameters were previously found in \cite{james2021evaluating} to be well described by the following behavioral model:
\begin{figure}
    \centering
    \includegraphics{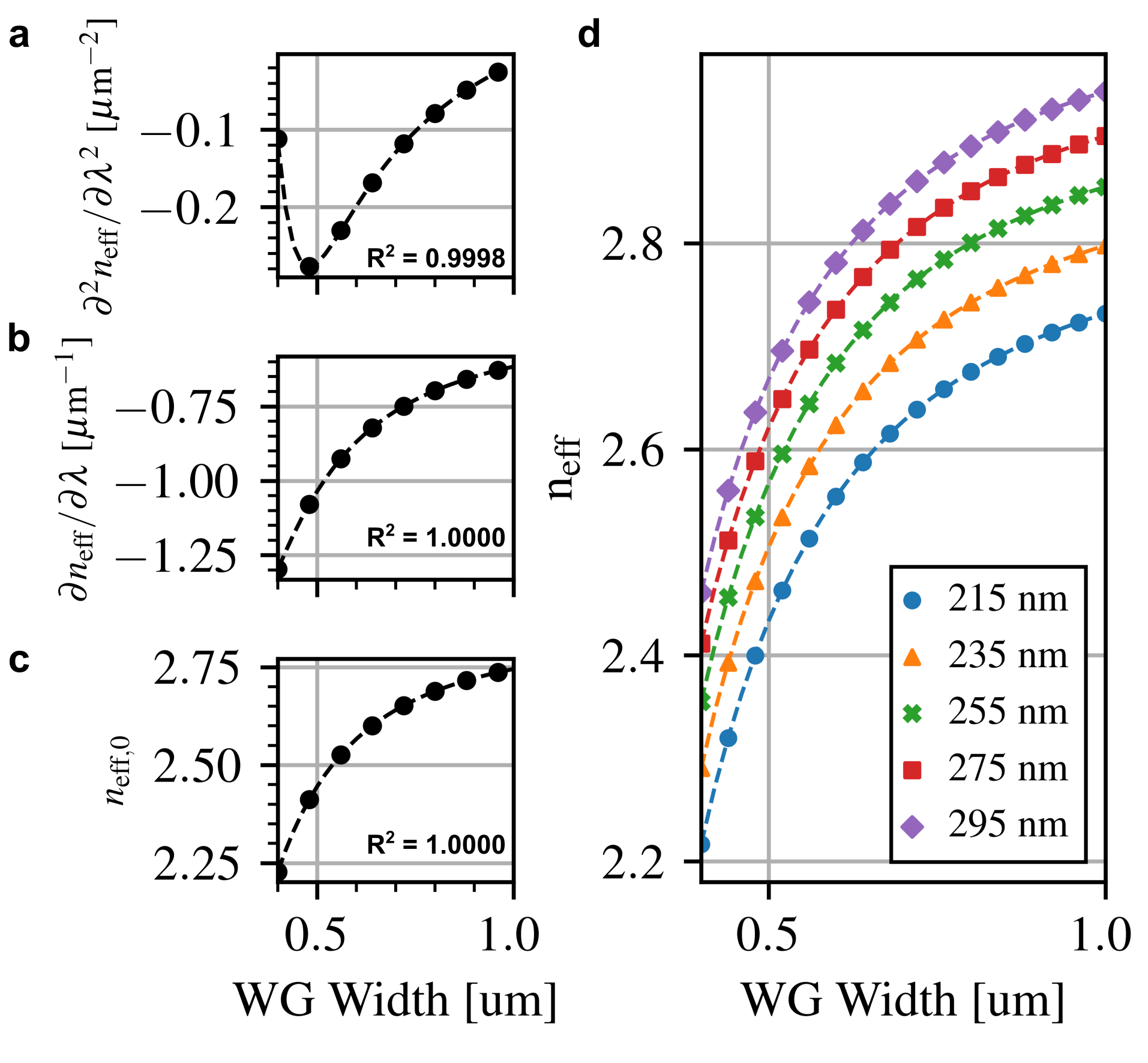}
    \caption{\textbf{a-c,} Plot of simulated (scatter) and modeled (dashed) $n_{\text{eff}}$ parameters $\left[\partial^2n_{\text{eff}}/\partial\lambda^2, \partial n_{\text{eff}}/\partial\lambda, n_{\text{\text{eff}}, 0}\right]$ vs waveguide width (respectively). These values were for a waveguide with a thickness of 220~nm at a wavelength of 1550~nm. \textbf{d,} Comparison of the model (dashed) and simulated (scatter) $n_{\text{eff}}$ vs waveguide width for different thicknesses. Simulated at 1550~nm.}
    \label{fig:fit_width_variation}
\end{figure}
\begin{equation}\label{eq:parameter_eq}
       \color{black}\frac{\partial^in_{\text{eff}}}{\partial\lambda^i}(w)=p_{i0}\cdot\frac{w^2+p_{i1}w+p_{i2}}{w^2+p_{i3}w+p_{i4}},
\end{equation}
for a total of fifteen model parameters. To verify correctness of the model, all three parameters were fitted \color{black}to the simulation data with \color{black} \eqref{eq:effective_index_model}-\eqref{eq:parameter_eq} using ordinary least squares (OLS) regression. The model was able to match all three parameters over the entire range of the width sweep (Fig. \ref{fig:fit_width_variation}a-c). The close matching of the modeled and extracted Taylor parameters means that our modification of \eqref{eq:parameter_eq} still preserves its ability to match the behavior of effective index as a function of wavelength. By extension, these three Taylor parameters allow for a robust description of \color{black}$n_{\text{eff}}$ \color{black} as a function of waveguide width (Fig. \ref{fig:fit_width_variation}d). The data also demonstrates this agreement is not unique to any particular waveguide thickness, with different thicknesses producing different sub-parameter fits.
\begin{figure}
    \centering
    \includegraphics{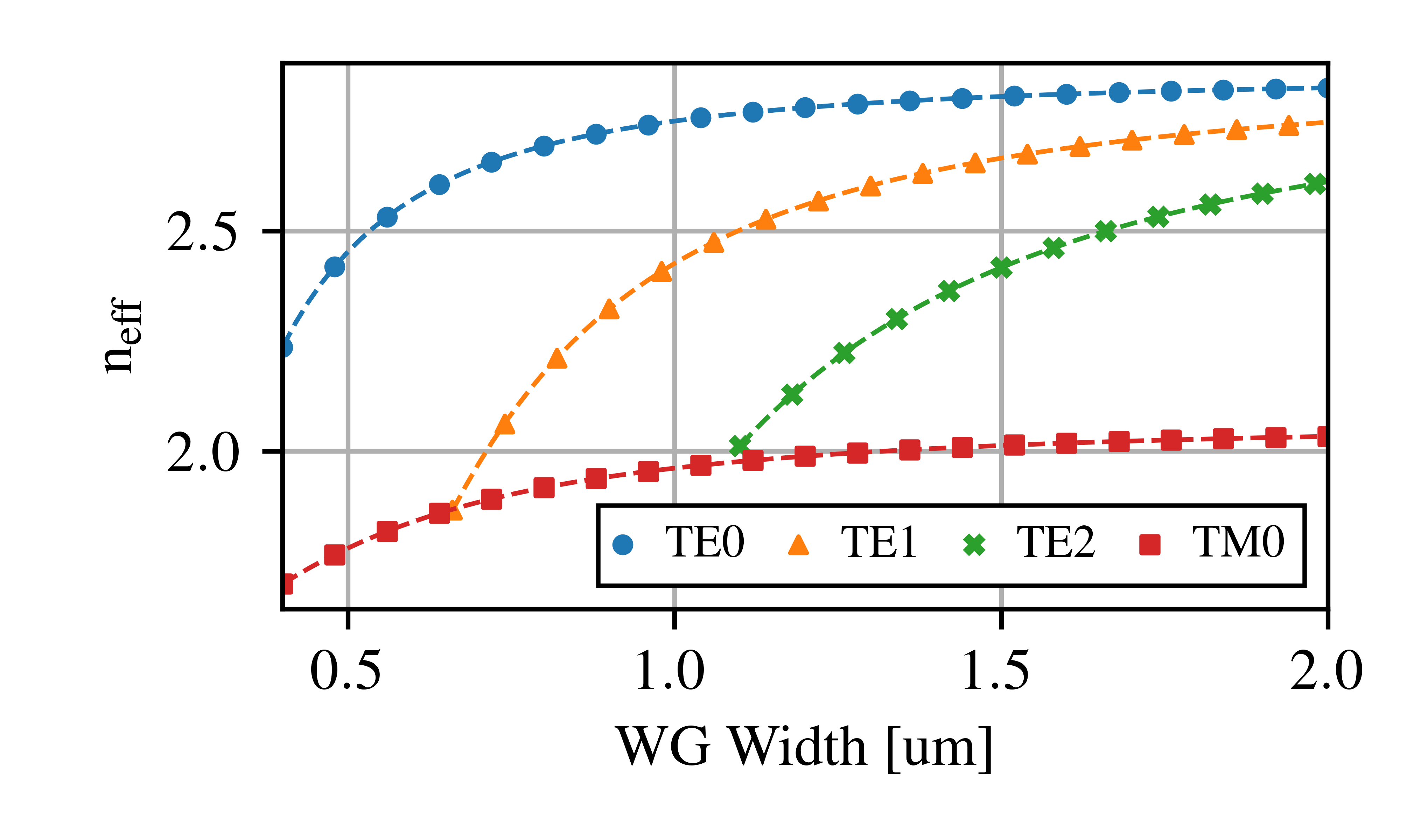}
    \color{black}\caption{Comparison between modeled (dashed) and simulated (scatter) $n_{\mathrm{eff}}$ for higher order modes and the fundamental TM mode. All waveguides were simulated with a thickness of 220~nm.}
    \label{fig:other_modes}
\end{figure}
Finally, it should be noted that both the numerator and denominator in \eqref{eq:parameter_eq} are polynomials of equal order. Our model consequently predicts that, for a given wavelength, the effective index will asymptotically approach a constant value as $w$ approaches infinity. The value that the model approaches as $w$ tends towards infinity can be interpreted as the equivalent \color{black}$n_{\text{eff}}$ \color{black} of an infinite slab of \color{black}the same thickness\color{black}:
\begin{equation}\label{eq:asymptote}
        \lim_{w \to \infty} n_{\text{eff}}(\lambda, w) = n_{\text{slab}}(\lambda).
\end{equation}
In this way, our behavioral model can elegantly capture all significant features of effective index for \color{black}the \color{black} design parameters of interest. \color{black}The model's accuracy holds true for higher order modes as well, provided that they are sufficiently far away from their respective waveguide cutoff condition (Fig. \ref{fig:other_modes}). \color{black}

\subsection{Scattering Loss}

\begin{figure}
    \centering
    \includegraphics{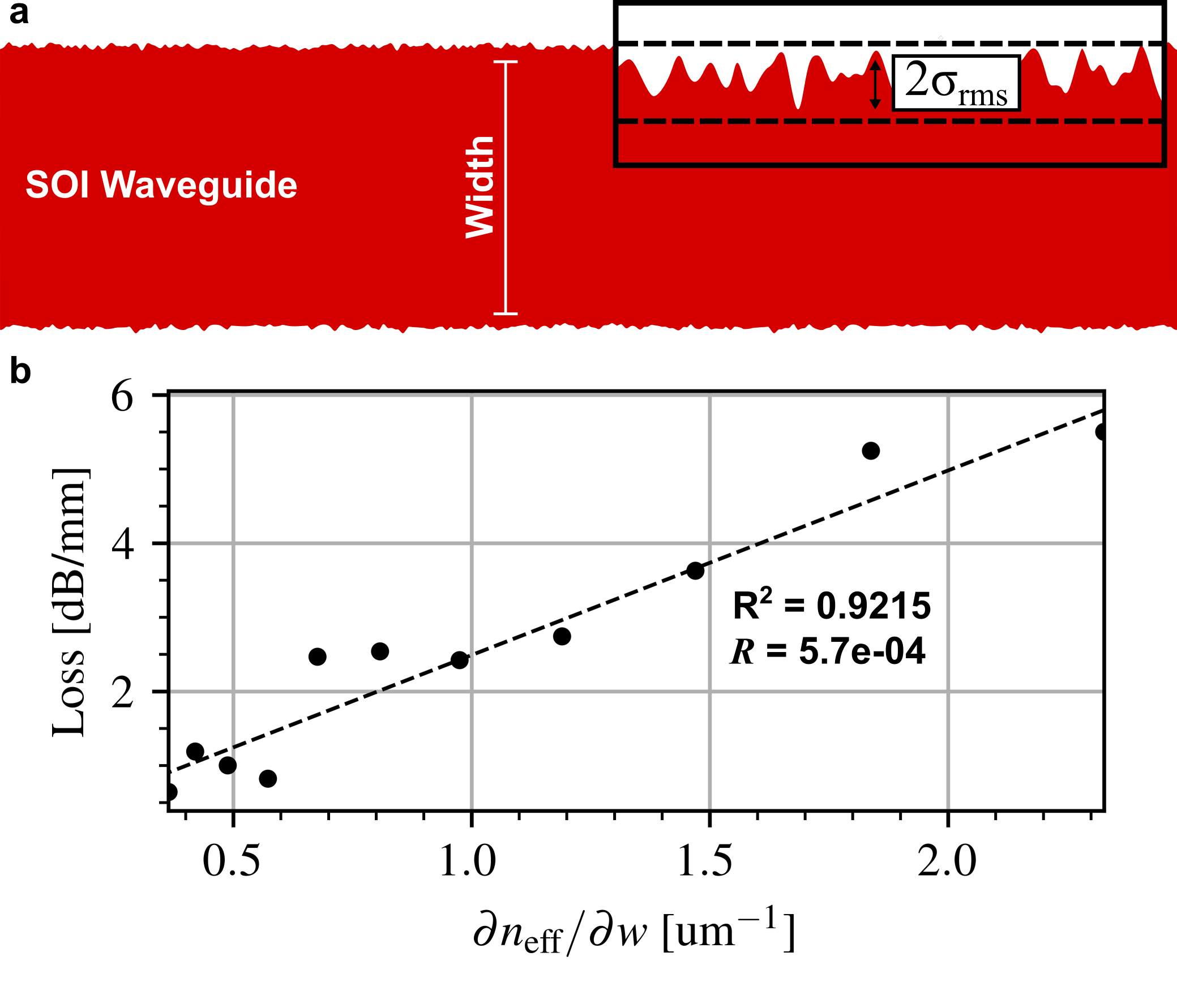}
    \caption{\textbf{a,} Graphical representation of a waveguide simulated with some sidewall roughness. The inset is a magnified view of the waveguide to clarify the definition of $\sigma_{\mathrm{rms}}$. \textbf{b,} Scattering losses estimated from FDTD compared to the fit using our model based on Lumerical MODE data.}
    \label{fig:scattering}
\end{figure}

Scattering loss due to sidewall roughness (SWR) \color{black} can be a significant \color{black} source of loss in most reported waveguide designs, making it critical for designers to accurately model \cite{heck2014ultra}. In this section, we demonstrate our model's ability to capture SWR loss as a function of waveguide geometry. It was first noted in \cite{melati2014real} that the traditional Payne and Lacey model of SWR-induced loss \cite{lacey1990radiation, payne1994theoretical} was found to be identical in behavior to the derivative of the effective index with respect to waveguide width:
\begin{equation}\label{eq:swr_loss_fit}
    \alpha_{\text{SWR}}(\lambda, w) = R\frac{\partial}{\partial w}\left[n_{\text{eff}}(\lambda, w)\right],
\end{equation}
where $R$ is a proportionality constant. As our model can describe \color{black}$n_{\text{eff}}$ \color{black} as a function of width, a closed-form representation of $\partial n_{\text{eff}}/\partial w$ can be exactly derived. This equation can then be fitted to measured waveguide loss data to extract the proportionality constant. We validate this by fitting \eqref{eq:swr_loss_fit} to the scattering loss of a 7~$\mu$m long SOI waveguide with some \color{black}SWR \color{black} wall roughness in  Lumerical 3D-FDTD (Fig. \ref{fig:scattering}a). The roughness Root Mean Square (RMS) and correlation length were arbitrarily chosen to be $\sigma_{\text{rms}}=5~\mathrm{nm}$ and $L_{\text{corr}}=1~\mathrm{\mu m}$ respectively. These parameters were then used to generate a random, anisotropic \color{black}SWR \color{black} on the waveguide walls \cite{jaberansary2013scattering}. Propagation losses were \color{black}simulated \color{black} for waveguide widths ranging from 450~nm to 850~nm. The results of the fitting are shown in Fig. \ref{fig:scattering}b, with our model closely matching trend of the scattering loss behavior extracted from FDTD simulations.
\subsection{Thermo-Optic Effect}\label{sec:thermooptic}

Our model can also completely describe the thermo-optic coefficient of an arbitrary waveguide geometry without the need for any thermal measurements. The thermo-optic coefficient of a waveguide mode most importantly requires knowledge of the confinement factor, which is the fraction of a mode's power confined within each constituent waveguide material. Kawakami showed in \cite{kawakami1975relation} that for a waveguide made up of N materials, each with with an index $n_k$ and a confinement factor $\Gamma_k$:
\begin{subequations}\label{eq:dispersion_confinement}
    \begin{align}
        \sum_k^N\Gamma_k n^2_k=n_{\mathrm{g}}n_{\mathrm{eff}}\\
        \sum_k\Gamma_k=1,\label{eq:power_conservation}
    \end{align}
\end{subequations}
where \eqref{eq:power_conservation} is derived from noting that the sum of all confinement factors must equal unity due to power conservation. A closed-form of the confinement factor for a two-material waveguide (e.g. SOI wires) can then be derived:
\begin{subequations}\label{eq:confinement_factor}
    \begin{align}
        \Gamma_{\text{core}}=\frac{\displaystyle n_{\mathrm{g}}n_{\mathrm{eff}} - n_{\text{clad}}^2}{\displaystyle n_{\text{core}}^2 - n_{\text{clad}}^2}\\
        \Gamma_{\text{clad}}=\frac{\displaystyle n_{\text{core}}^2 - n_{\mathrm{g}}n_{\mathrm{eff}}}{\displaystyle n_{\text{core}}^2 - n_{\text{clad}}^2},
    \end{align}
\end{subequations}
where $\Gamma_{\text{core}}$ is the power contained in the waveguide core and $\Gamma_{\text{clad}}$ is the power contained in the cladding.

Next, we must obtain an expression that describes the thermo-optic effect on \color{black}$n_{\text{eff}}$ \color{black}  in terms of the confinement factor. A common approximation of the thermo-optic coefficient of \color{black}$n_{\text{eff}}$ \color{black}  is
\begin{equation}\label{eq:wrong_thermal_equation}
    \frac{\partial n_{\text{eff}}}{\partial T} \approx \Gamma_1\frac{\partial n_{\text{1}}}{\partial T} + \Gamma_2\frac{\partial n_{\text{2}}}{\partial T} + \dots,
\end{equation}
where $\delta$ represents a small perturbation in the values, $\Gamma_n$ is the confinement of the mode within material $n$ and $\partial n_{\text{n}}/\partial T$ is the thermo-optic coefficient of material $n$ \cite{winnie2008polymer}. Though this equation is widely used \cite{jean2021sulfur,zhang2016chip,zhou2015lowering} and may be accurate in certain scenarios, to the authors' knowledge it has never been demonstrated to be a generally accurate approximation. We therefore start from first principles and consider a general perturbation of the wave equation \cite{yariv2007photonics}:
\begin{equation}\label{eq:wave_equation_perturbed}
    \delta\left[\beta^2_{\mathrm{eff}}\right]=\Gamma_{\text{core}}\frac{\omega^2}{c^2}\delta\left[n_{\text{core}}^2\right]+\Gamma_{\text{clad}}\frac{\omega^2}{c^2}\delta\left[n_{\text{clad}}^2\right],
\end{equation}
where $\beta_{\mathrm{eff}}$ is the effective wavenumber, $\Gamma_{\text{core}}$ is the confinement in the waveguide core, $\Gamma_{\text{clad}}$ is the confinement in the waveguide cladding, and $n_{\text{core}}$ and $n_{\text{clad}}$ are the core and cladding indices respectively. Carrying this operation through and combining with \eqref{eq:effective_index_model} (see Appendix \ref{app:thermal} for details) yields:
\begin{subequations}\label{eq:thermooptic_effect}
        \begin{align}
        n_{\text{eff}}(\lambda,w,T)\approx n_{\text{eff}, T_0}(\lambda,w) + \frac{\partial n_{\text{eff}}}{\partial T}\left(T-T_0\right)\\
        \frac{\partial n_{\text{eff}}}{\partial T}=\Gamma_{\text{core}}\frac{n_{\text{core}}}{n_{\text{eff, T$_0$}}}\frac{\partial n_{\text{core}}}{\partial T}+\Gamma_{\text{clad}}\frac{n_{\text{clad}}}{n_{\text{eff, T$_0$}}}\frac{\partial n_{\text{clad}}}{\partial T},
        \end{align}
\end{subequations}
where $n_{\text{eff}, T_0}$ is the \color{black}$n_{\text{eff}}$ \color{black}  at some reference temperature $T_{0}$. \color{black} The key addition to \eqref{eq:thermooptic_effect} compared to prior literature is the scaling of each thermo-optic term by ratio between the material and effective indices. As the index contrast between the core and cladding decreases, our model will approach the \eqref{eq:wrong_thermal_equation}. Thus it is clear that our model will outperform \eqref{eq:wrong_thermal_equation} in accuracy when describing high index contrast materials, such as the SOI waveguide geometries prevalent in SiPh.\color{black}

\begin{figure}
    \centering
    \includegraphics{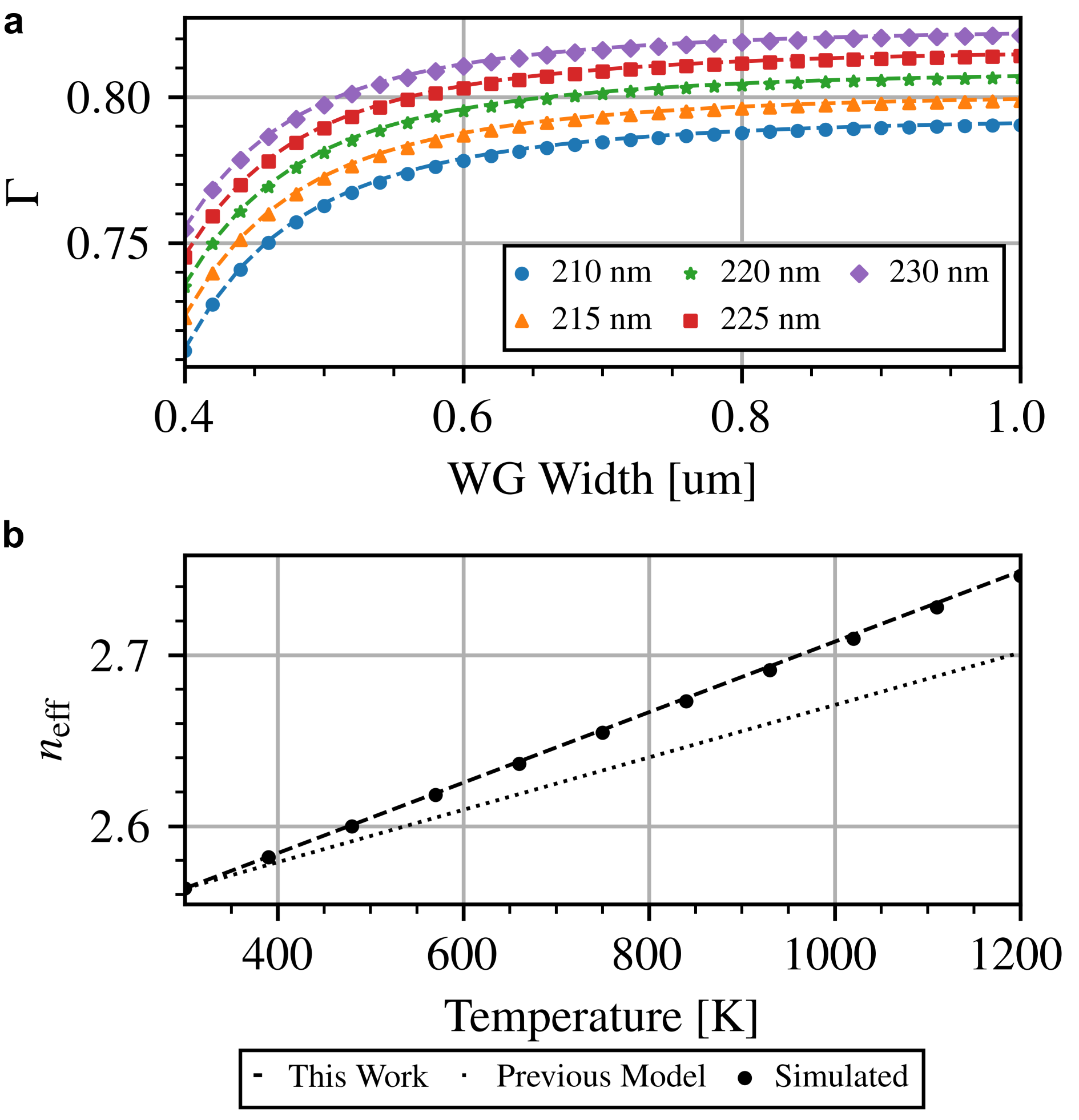}
    \caption{\textbf{a} Modeled (dashed) and simulated (scatter) confinement factor vs waveguide width for different thicknesses. \color{black}\textbf{b,} Comparison between simulated (scatter), previously reported model (dotted, Eq. \eqref{eq:wrong_thermal_equation}) and our work (dashed line, Eq. \eqref{eq:thermooptic_effect}) describing $n_{\text{eff}}$ vs Temperature of a 480~x~220~nm waveguide.}
    \label{fig:neff_thermal}
\end{figure}

With these expressions, our confinement factor and the thermo-optic coefficient models can be validated. The simulated confinement factor is compared to our model prediction at 1550~nm in Fig. \ref{fig:neff_thermal}a. The optical properties of silicon and silicon dioxide used in our model were taken directly from \cite{palik1998handbook}. There was a near perfect agreement between the modeled and simulated confinement factor, showing that the general behavior of confinement factor is captured by our model (Fig. \ref{fig:neff_thermal}a). The modeled thermo-optic coefficient is validated by simulating how the \color{black}$n_{\text{eff}}$ \color{black} of a SOI waveguide varies with temperature using Lumerical MODE (Fig. \ref{fig:neff_thermal}b). Silicon was assumed to have a thermo-optic coefficient of \num{1.9e-4}~K$^{-1}$ \cite{Frey_2006} and SiO$_2$ was assumed to have a thermo-optic coefficient of \num{1e-5}~K$^{-1}$ \cite{elshaari2016thermo}. The model and simulations show exceptional agreement from 300~-~1200~K, despite the fact that our model does not require any data from thermal simulations or measurements. \color{black} As predicted, the previously reported model of the thermo-optic effect \eqref{eq:wrong_thermal_equation} significantly under-predicts the expected change in $n_{\mathrm{eff}}$. \color{black} It should be noted that in real devices, waveguide geometry itself is a function of $T$ due to thermal expansion. This can be accounted for by modeling $w$ as a function of $T$. Experimental results in Section \ref{sec:thermo_validation}, however, show that assuming a constant width geometry provides sufficient accuracy.

\color{black} Having a model of the thermo-optic effect that is accurate over a wide range of conditions like this one holds a great \color{black}deal of potential to enable more robust design exploration, such as evaluating photonic waveguide heater designs \cite{coenen2022thermal, jacques2019optimization}, characterizing self-heating in micro-resonators \cite{deCea:19}, or studying the effect of ambient temperature fluctuations in a system.

\color{black}\subsection{Parameter Extraction}\label{sec:extraction}\color{black}
The practical utility of a compact model is greatly determined by the associated parameter extraction procedure to connect the model to a given foundry process. This is particularly important when developing statistical models, as accurate parameter extraction is the only way to guarantee that process variations are accurately reflected in the model. A popular solution is to leverage the phase-sensitivity of interferometric optical filters---such as Mach-Zehnder interferometers (MZIs), microresonators, or arrayed waveguide gratings (AWGs)---to monitor process variations across a wafer. Regardless of the chosen device, a shared difficulty lies in accurately guessing what particular interference fringe position corresponds to a particular fringe order \cite{oton2016silicon, lu2017performance, xing2018accurate}. \color{black}Our method is based on the curve-fitting method presented in \cite{xing2018accurate} and \cite{Deng:16}, with some additional steps described to include waveguide dispersion as an extracted parameter.\color{black}

The first step in parameter extraction is to characterize the group index \color{black}($n_g$) \color{black}of a fabricated interferometer from a \color{black}wavelength \color{black} sweep of the device. To enable this, \eqref{eq:effective_index_model} is rearranged into a more suitable form:
\begin{subequations}\label{eq:reformulated_index_model}
    \begin{align}
        n_{\text{eff}}(\lambda)=\frac{1}{2}\frac{\partial^2n_{\text{eff}}}{\partial\lambda^2}\lambda^2+B\lambda+C\label{eq:B_derivation}\\
        B = \frac{\partial n_{\text{eff}}}{\partial\lambda} - \frac{\partial^2n_{\text{eff}}}{\partial\lambda^2}\lambda_0\label{eq:B_def}\\
        C = \frac{1}{2}\frac{\partial^2n_{\text{eff}}}{\partial\lambda^2}\lambda^2_0 - \frac{\partial n_{\text{eff}}}{\partial\lambda}\lambda_0 + n_{\text{eff,0}},\label{eq:C_def}
    \end{align}
\end{subequations}
where $B$ and $C$ are fitting parameters that aggregate the 1$^{\text{st}}$ and 0$^{\text{th}}$ order terms from \eqref{eq:effective_index_model} respectively. Following the procedure described in \cite{Deng:16}, it is first noted that the fringe condition of \color{black}an \color{black}inteferometric device is described by
\begin{equation}\label{eq:resonance_condition}
    \phi=\frac{2\pi}{\lambda}n_{\text{eff}}(\lambda)L=2\pi m,
\end{equation}
where $\phi$ is the phase difference between the interferometry arms, $L$ is the path length of the interferometer, $\lambda$ is a particular fringe wavelength, and $m$ is an integer corresponding to the particular fringe order. To extract our model parameters, a \color{black}wavelength \color{black}sweep of the interferometric device is required. Once this is performed, a peak finding algorithm can be used to detect the wavelength of all detected fringes. A function that relates the relative fringe locations to the \color{black}$n_g$ \color{black} of the waveguide is now required. This can be done by defining a continuous function that will yield an integer value at each of the detected fringe locations. Let $m_0$ represent the particular fringe order corresponding to an arbitrarily chosen reference fringe located at $\lambda_0$. The fringe order $m$ of any other fringe can be defined relative to this reference as
\begin{equation}\label{eq:deng_m_equation}
    m = m_0 + \int_{\lambda_0}^{\lambda}\frac{dm}{d\lambda}\text{d}\lambda=m_0+n_\text{g}L\cdot\left(\frac{1}{\lambda}-\frac{1}{\lambda_0}\right).
\end{equation}
This continuous function now allows us to redefine the measured fringes into a form suitable for parameter extraction. A reference fringe variable $n$ is now defined by letting $m=(m_0+n)$. Inserting this back into \eqref{eq:deng_m_equation} produces:
\begin{equation}\label{eq:deng_n_equation}
    n=n_g L\cdot\left(\frac{1}{\lambda_n}-\frac{1}{\lambda_0}\right),
\end{equation}
where each relative fringe $n$ is located at an associated wavelength $\lambda_n$. Using \eqref{eq:deng_n_equation}, the \color{black}$n_g$ \color{black} of the measured device is now directly related to the measured fringe locations. This fitting equation must now be extended to our specific model parameters. The \color{black}$n_g$ \color{black} of a waveguide is defined to be
\begin{equation}\label{eq:group_index}
    n_g=n_{\text{eff}}-\lambda\frac{\partial n_{\text{eff}}}{\partial\lambda}.
\end{equation}
Combining with \eqref{eq:B_derivation} yields an expression for \color{black}$n_g$ \color{black} in terms of our compact model:
\begin{equation}\label{eq:group_index_compact_model}
    n_{\text{g}}= C - \frac{1}{2}\frac{\partial^2n_{\text{eff}}}{\partial\lambda^2}\lambda^2.
\end{equation}
By inserting \eqref{eq:group_index_compact_model} back into \eqref{eq:deng_n_equation}, we can derive an OLS regression-compatible expression:
\begin{subequations}\label{eq:group_regression_equations}
    \begin{align}
        n=C\Lambda_C - \frac{\partial^2n_{\text{eff}}}{\partial\lambda^2}\Lambda_S\label{eq:reg_a}\\
        \Lambda_C=L\cdot\left(\frac{1}{\lambda_n}-\frac{1}{\lambda_0}\right)\label{eq:reg_b}\\
        \Lambda_S=\frac{L}{2}\cdot\left(\lambda_n-\frac{\lambda_{n}^{2}}{\lambda_0}\right)\label{eq:reg_c},
    \end{align}
\end{subequations}
where $[\Lambda_C, \Lambda_S]$ are explanatory variables. Performing an OLS regression between $n$ and $[\Lambda_C, \Lambda_S]$ gives us two of our three fitting parameters in \eqref{eq:reformulated_index_model}. Finally, $B$ can be calculated by combining equations \eqref{eq:resonance_condition} and \eqref{eq:B_derivation}:
\begin{equation}\label{eq:B_from_measurements}
    B=\frac{m}{L} - \frac{1}{2}\frac{\partial^2n_{\text{eff}}}{\partial\lambda^2}\lambda_m - \frac{C}{\lambda_m},
\end{equation}
where the only uncertainty is what fringe order $m$ corresponds to each measured fringe $\lambda_m$. Once $B$ is determined from \eqref{eq:B_from_measurements}, \eqref{eq:B_def} and \eqref{eq:C_def} can be used to determine the original fitting parameters in \eqref{eq:effective_index_model}. It should be noted that each detected fringe ($m$, $\lambda_m$) location will yield very small variations in the B value due to resolution-based uncertainty in the exact value for $\lambda_m$. For a best guess, all values $B_m$ taken from each measured fringe $\lambda_m$ should be averaged together.

\begin{figure}
    \centering
    \includegraphics{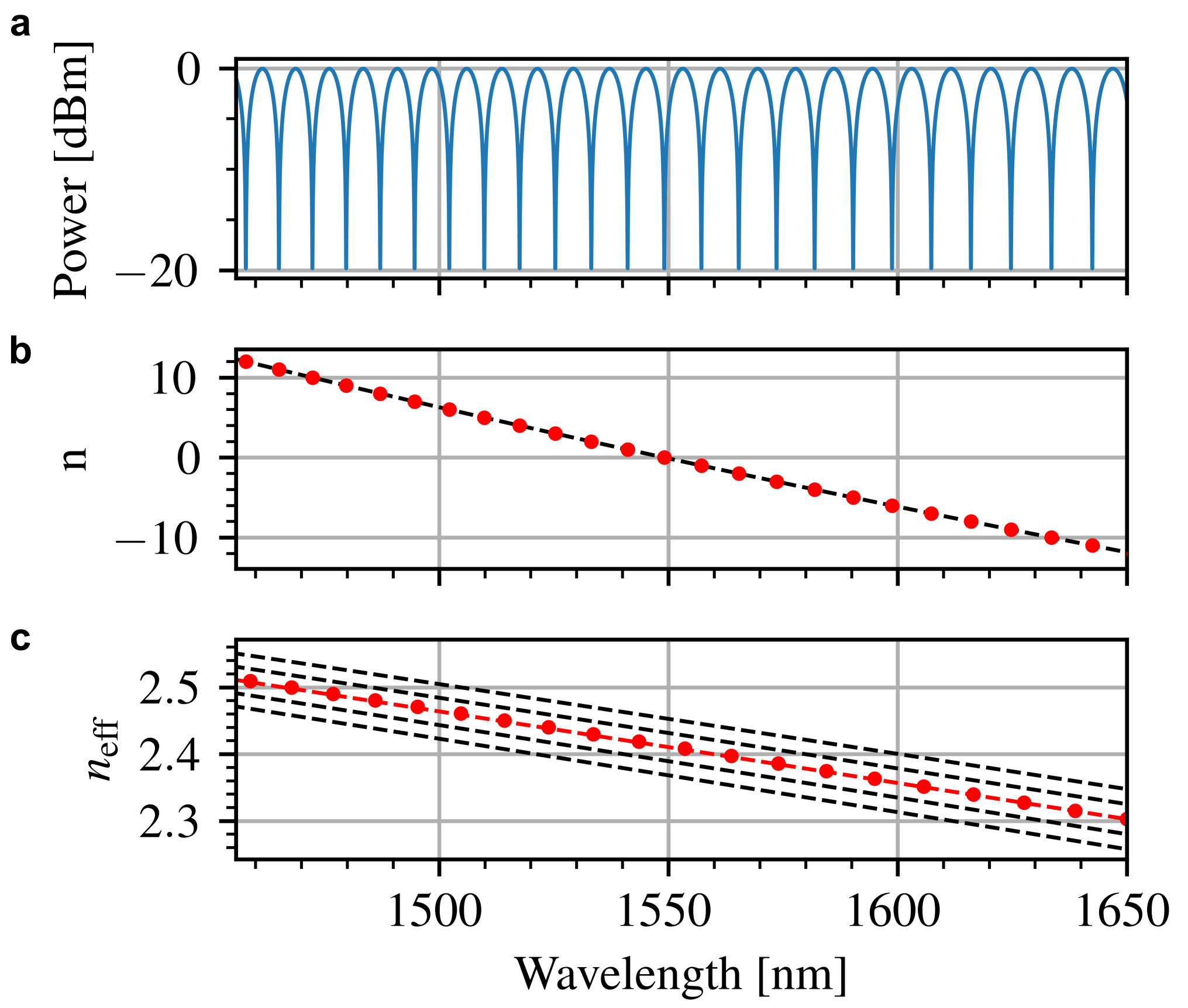}
    \color{black}\caption{\textbf{a,} \color{black}Captured spectrum of simulated MZI used for parameter extraction. The waveguide mode was simulated in Lumerical MODE, and then exported to a MZI waveguide simulation block in Lumerical INTERCONNECT. \color{black} \textbf{b,} Linear Regression of fringe wavelengths to extract the \color{black}$n_g$ \color{black} performed on the detected fringes from \textbf{a}. \textbf{c,} Possible \color{black}$n_{\text{eff}}$ \color{black} solutions (black, dashed), along with the actual solution (red), determined by the \color{black}$n_g$ \color{black} extracted in \textbf{b}.}
    \label{fig:fig_simulated_ring_fit_data}
\end{figure}

To validate this method under ideal conditions, an MZI constructed using 480~nm x 220~nm waveguides is simulated in Lumerical INTERCONNECT. To ensure accuracy, the waveguide's \color{black}$n_{\text{eff}}$ \color{black} was first simulated in MODE and then exported to a MODE Waveguide element in INTERCONNECT. As the full-width half-maximum (FWHM) of the MZI does not affect the  extracted \color{black}$n_{\text{eff}}$\color{black}, the waveguides were arbitrarily assumed to have a 2.5 dB/cm loss and the coupling coefficient was chosen to ensure critical coupling. The spectrum of the simulated MZI is shown in Fig. \ref{fig:fig_simulated_ring_fit_data}a. Fringe locations were extracted using a peak finding algorithm. The fringe located closest to the center of the sweep was arbitrarily chosen as $n=0$. Using \eqref{eq:group_regression_equations}, OLS regression found $\partial^2n_{\text{eff}}/\partial\lambda^2 = -0.136$ $\mu$m$^{-2}$ and $C = 3.9215$ (Fig. \ref{fig:fig_simulated_ring_fit_data}b). From here, the family of solutions for $n_{\text{eff}}$ is plotted in Fig. \ref{fig:fig_simulated_ring_fit_data}c. Each particular solution corresponds to a different guess on the fringe orders detected, e.g. $m_0 = 52$ vs. $m_0 = 53$. The separation between each \color{black}$n_{\text{eff}}$ \color{black} solution plotted in \ref{fig:fig_simulated_ring_fit_data}c is determined by the free-spectral range (FSR) of the interferometer, with a larger FSR corresponding more widely separated solutions.

To determine the correct fringe order of the reference we use the fact that, from the simulations performed in Section \ref{sec:model}, we know the waveguide geometry has an \color{black}$n_{\text{eff}}$ \color{black} of 2.411 at the reference fringe location. In Section~\ref{sec:model_variability_discussion} we explain how to increase the accuracy of this estimation to avoid errors introduced by this simulation. From this, the reference fringe order is found to be $m_0\approx114.03$. Since fringe orders must be integer numbers, the result is rounded to the nearest integer 114. By combining \eqref{eq:B_derivation}-\eqref{eq:C_def}, the original fitting coefficients are found to be $\partial n_{\text{eff}}/\partial\lambda = -1.078$ $\mu$m$^{-1}$ and $n_{\text{eff,0}} = 2.411$. To evaluate accuracy of our extraction, we define the relative error between the extracted and simulated \color{black}$n_{\text{eff}}$'s \color{black}$\sigma_{\text{error}}$ by:
\begin{equation}\label{eq:error_eq}
    \sigma_{\text{error}}=\sqrt{\frac{\int\left(n_{\text{eff, model}} - n_{\text{eff, sim}}\right)^2 \text{d}\lambda}{\int n_{\text{eff, sim}}^2 \text{d}\lambda}},
\end{equation}
\color{black}where $n_{\text{eff, sim}}$ is the effective index from the MODE simulation, used as a reference to quantify our method's accuracy, and $n_{\text{eff, model}}$ is the result from applying our extraction method to the simulated MZI. \color{black}Upon evaluation, the total relative error was found to be 0.017\%. Since the order of the reference fringe is correct, the remaining model error is attributed to inaccuracies in the initial regression fit using \eqref{eq:reg_a}-\eqref{eq:reg_c}. 

\color{black}\section{More Robust $n_{\mathrm{eff}}$ Extraction Under Process Variability}\color{black}\label{sec:model_variability_discussion}
\begin{figure}
    \centering
    \includegraphics{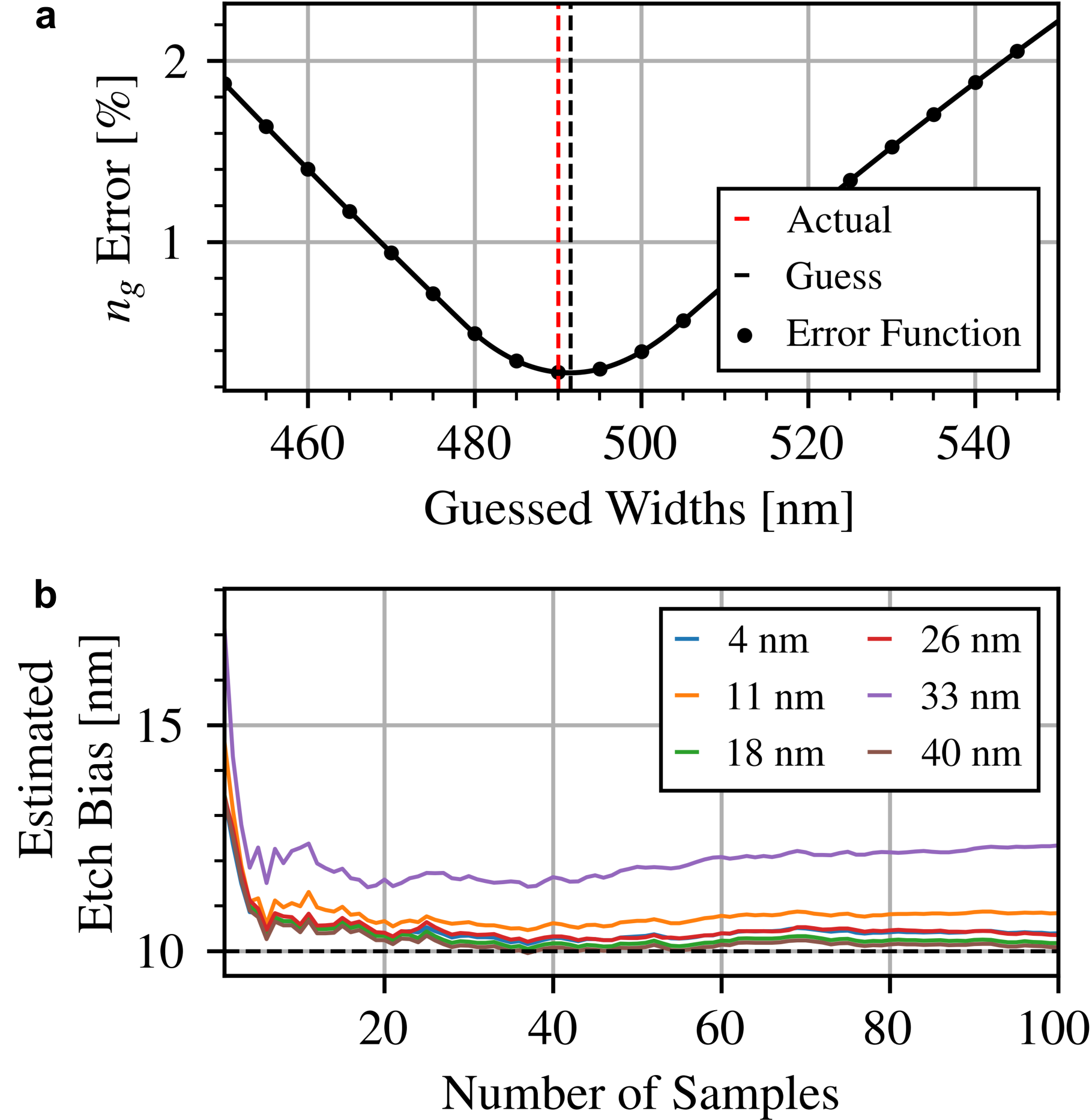}
    \color{black}\caption{\textbf{a,} Plot of the \color{black}$n_g$ \color{black} error function \color{black}for one sample. \color{black}The error function shows a minimum at roughly 491.5~nm, which closely agrees with the actual waveguide width of 490~nm. \textbf{b,} Convergence of the etch bias estimate for different numbers of samples averaged.}
    \label{fig:ng_error_plot}
\end{figure}  

The reliability of the extraction is highly sensitive to the guessed value of the reference fringe order. For the example in Section \ref{sec:extraction}, we used \textit{a~priori} knowledge of the \color{black}$n_{\text{eff}}$ \color{black} at the reference fringe to estimate its order. \color{black} Therefore, \color{black}any deviation between the assumed and actual waveguide dimensions risks introducing error. By noting that the initial order estimate rounded to the nearest integer, we can use \eqref{eq:resonance_condition} to define a boundary beyond which our fringe order guess will be incorrect 
\cite{xing2018accurate}: 
\begin{equation}\label{eq:resonance_order_constraint}
    |\Delta m|=|n_{\text{eff, actual}} - n_{\text{eff, guess}}|\leq\frac{\lambda_{m_0}}{2L}.
\end{equation}
We can see that, to raise confidence in the guessed fringe order, either the accuracy of our \color{black}$n_{\text{eff}}$ \color{black} guess must be increased or the interferometric path length must be decreased. As explained in Section \ref{sec:extraction}, our extraction method begins by directly extracting the \color{black}$n_g$ \color{black} of a given interferometer via optical sweep. Process variations will therefore appear as variations in the extracted values for $\partial^2n_{\text{eff}}/\partial\lambda^2$ and $C$. By measuring several devices of the same drawn width across the all measured dies, wafers, and lots, the influence of the random width and thickness variations can be eliminated by averaging their extracted fitting parameters. As the sample size becomes sufficiently large---with the necessary sample size being a function of the severity of the process variations---any remaining deviation between the nominal and averaged parameters will be the result of a systemic etch biases on the waveguide width. We therefore propose estimating this etch bias by creating a preliminary \color{black}$n_{\text{eff}}$ \color{black} model based on the results of a photonic mode solver, such as Lumerical MODE. Using this model, an equivalent waveguide width can be found by minimizing the error function
\begin{equation}\label{eq:error_function}
    \min_{w}\sqrt{\frac{\int\left[n_{\text{g, model}}(w, \lambda) - n_{\text{g, meas}}(\lambda)\right]^2 \text{d}\lambda}{\int n_{\text{g, meas}}^2(\lambda) \text{d}\lambda}},
\end{equation}
where $n_{\text{g, meas}}$ is the extracted model of \color{black}$n_g$ \color{black} using the averaged extracted parameters and $n_{\text{g, model}}$ is the simulation-based, width-dependent \textit{a~priori} model of \color{black}$n_{\text{eff}}$. \color{black} The \color{black}$n_{\text{eff}}$ \color{black} of our equivalent waveguide width can then be plugged into the \textit{a~priori} model to provide a more accurate fringe order estimate. In this way, we can increase the accuracy of our guessed effective index\color{black}, regardless of whether the modeled waveguide composition is accurate to the virtual device composition. 

We now discuss the robustness of this optimization routine in the presence of other systemic non-idealities and its ability to perform etch bias correction. To do this, we need a 'ground truth' value for $n_{\mathrm{eff}}$, which we obtain by simulating all the  non-idealities in Lumerical MODE. Subsequently we perform  the parameter extraction using Lumerical INTERCONNECT. By comparing the extracted $n_{\mathrm{eff}}$ to the known simulated value for $n_{\mathrm{eff}}$, we can directly evaluate the robustness of our methodology.\color{black}

\subsection{Statistical Geometric Variation}
To test the extraction procedure's accuracy under process variations, a simulation of 100 random variations on the waveguide geometry was run. The nominal waveguide dimensions were assumed to be 480~x~220~nm. To simulate systemic variations, each waveguide was arbitrarily assumed to have an etch bias of +10~nm. Random fluctuations were simulated by subjecting each device to a normally distributed variation of $3\sigma = 5~\mathrm{nm}$ on both the waveguide width and thickness, as this value is consistent with the worst-case reported values for geometric variations \cite{lu2017performance, xing2018accurate, zortman2010silicon}. Each mode profile was then exported to INTERCONNECT and simulated with interferometer FSRs ranging from 4 - 40~nm to investigate the effect this had on the extraction error. The resulting error function for one of these samples, with a ground truth width of 490nm, is shown in Fig. \ref{fig:ng_error_plot}a. We see the convergence behavior of the etch bias estimate evolves as a function of device sample size increases for several FSR designs in Fig. \ref{fig:ng_error_plot}b. It can be seen that all FSR designs can yield at least an estimated etch bias within 2 nm of the actual value, indicating the utility of our etch bias correction. 
\begin{figure}
    \centering
    \includegraphics{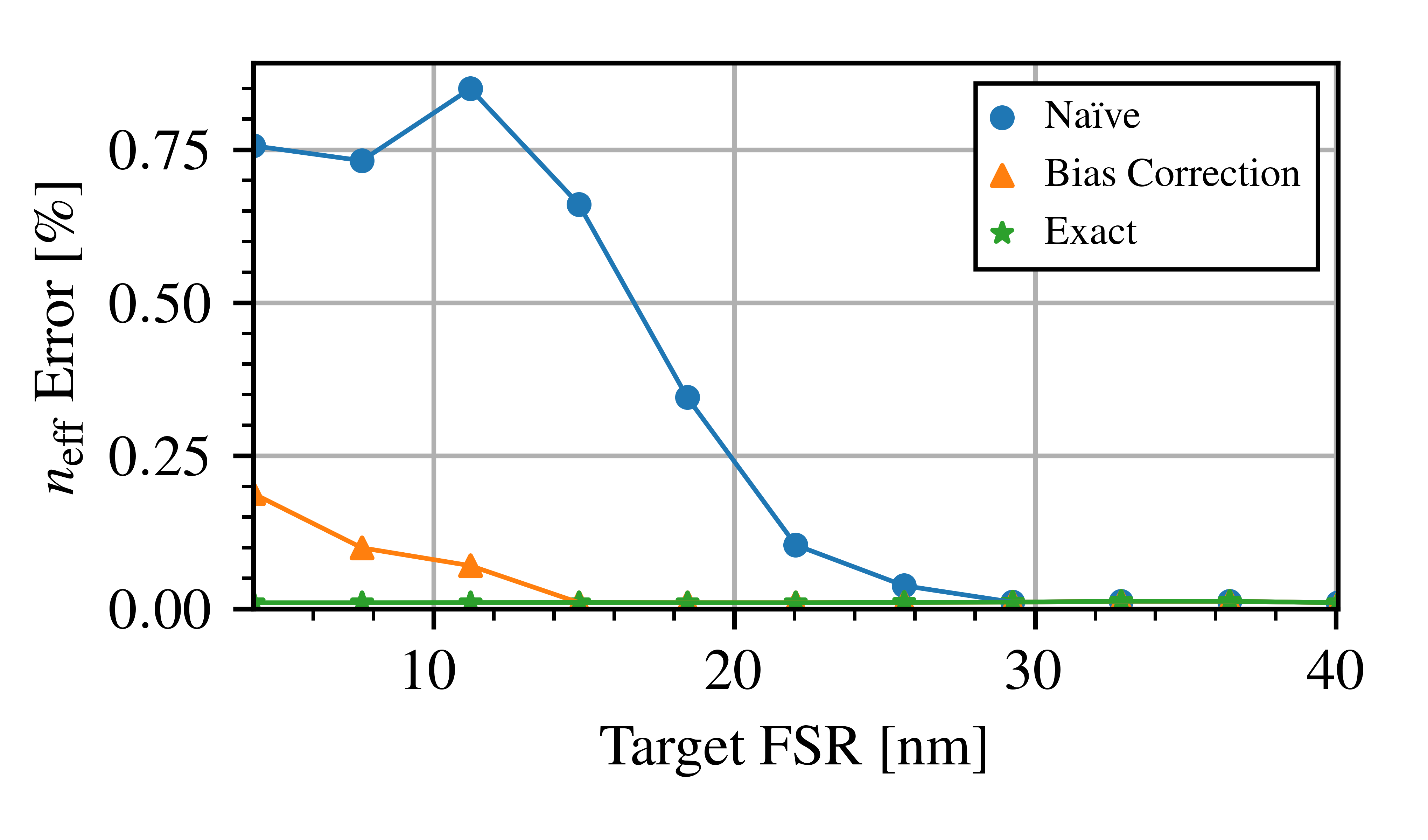}
    \color{black}\caption{Plot of mean error in \color{black}$n_{\text{eff}}$ \color{black} over the simulation bandwidth per simulated device. Each FSR was simulated with 100 random deviations from the target waveguide geometries. Both width and thickness were assumed to have a $3\sigma = 5~\mathrm{nm}$.}
    \label{fig:FSRvsError}
\end{figure}

Fig. \ref{fig:FSRvsError} shows the relationship between the average, per sample error and the interferometer FSR. The error is measured in three scenarios: i.) a `na\"ive' case, where the fringe order is estimated assuming no etch bias; ii.) where the fringe order is estimated through our etch bias prediction methodology, based on 30 measured samples; and iii.) where the exact \color{black}$n_{\text{eff}}$ \color{black} from simulations is used to determine the actual fringe orders. The last scenario, that produced an average per sample error of roughly 0.017\% represents an error floor for the first two. This error floor is completely determined by errors in the initial \color{black}$n_g$ \color{black} regression, as well as any fundamental limitations in our chosen behavioral model. As the FSR is increased, the average per sample error in both cases improves steadily until it reaches the aforementioned floor. This is consistent with \eqref{eq:resonance_order_constraint}, indicating that a larger FSR corresponds to a wider margin of error for the fringe order estimate. For both the na\"ive and bias compensation methods, there is a critical FSR value beyond which the fringe order is correctly estimated for all samples. It is clear, however, that estimating the presence of any etch biases drastically improves the fringe order accuracy, reaching the error floor for a much smaller FSR than when using the na\"ive method.

\begin{figure}
    \centering
    \includegraphics{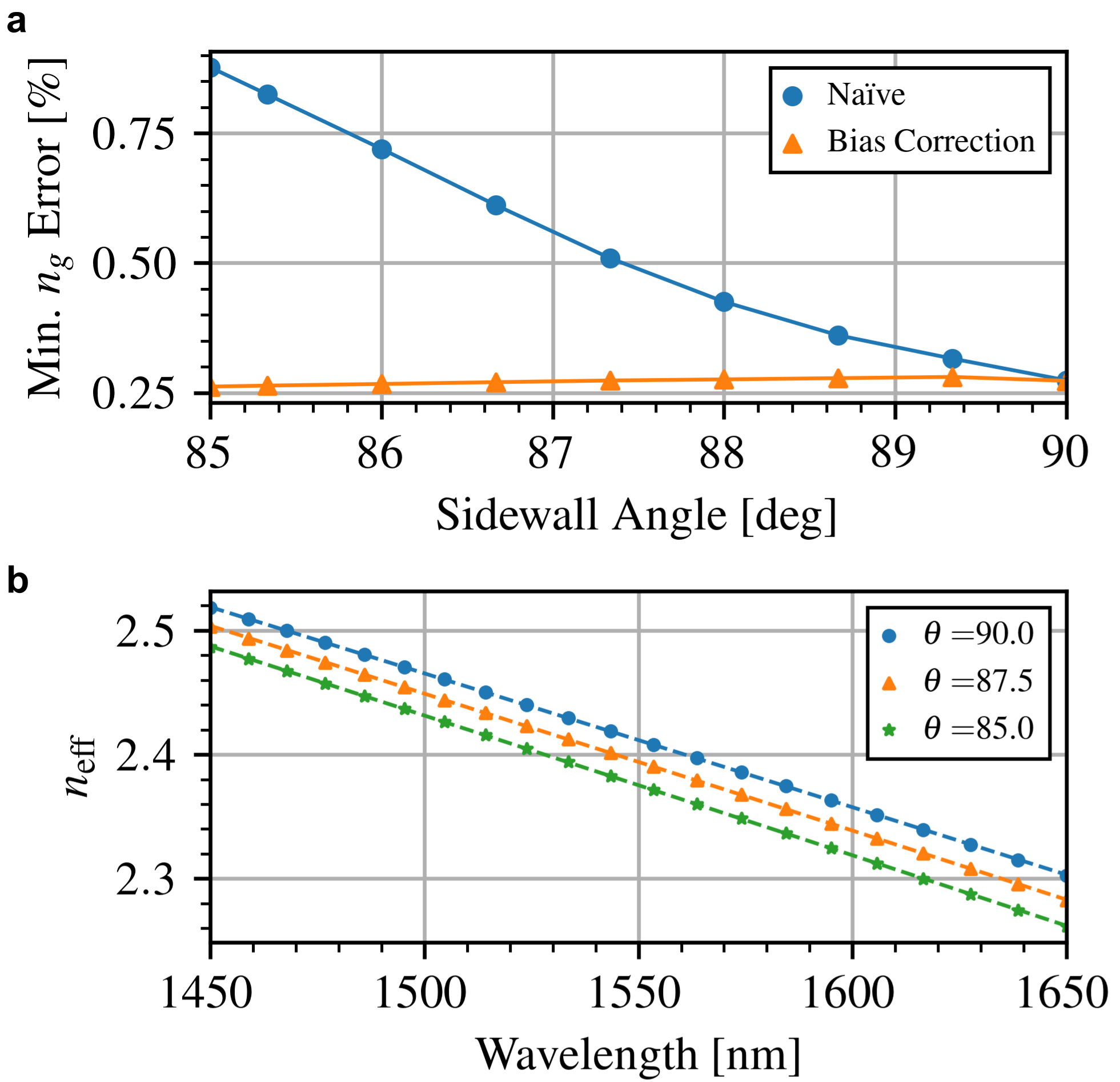}
    \color{black}\caption{\textbf{a,} \color{black}$n_g$ \color{black} relative error vs simulated sidewall angle. \textbf{b,} Comparison between the simulated (scatter) and estimated (dashed) \color{black}$n_{\text{eff}}$ \color{black} for different sidewall angles.}
    \label{fig:sidewall_correction}
\end{figure}
\begin{figure*}
    \centering
    \includegraphics{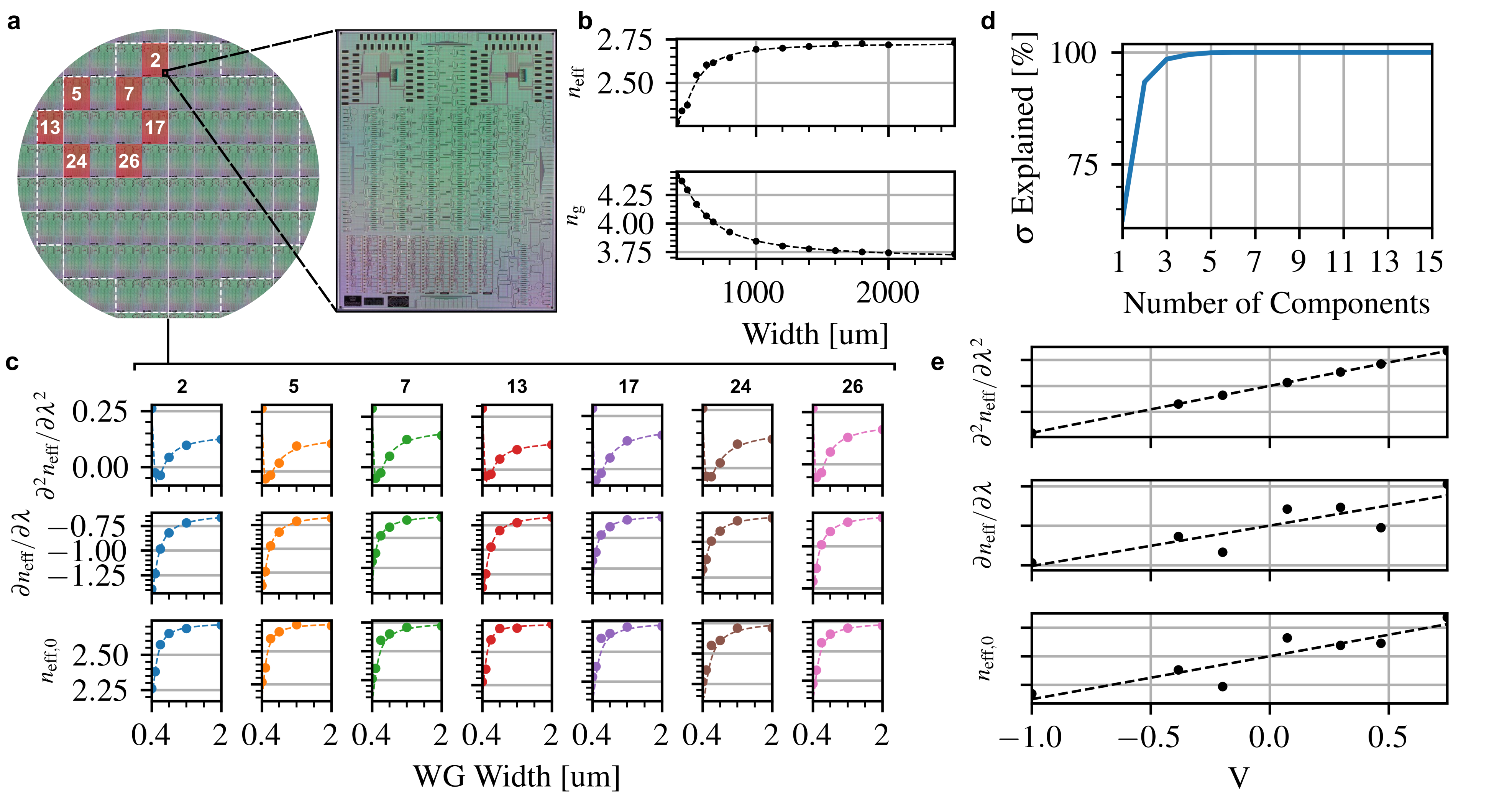}
    \color{black}\caption{\textbf{a,}  Illustration of measured reticles on a custom 300 mm wafer, with a blown-up microscopic
image of a die with 135 MZIs. \textbf{b,} Nominal $n_{\mathrm{eff}}$ and $n_g$ model extracted from device measurements. \textbf{c,} Width-based model extraction for each die tested. \textbf{d,} Total model parameter $\mu$m variance $\sigma$ explained vs number of principal components included.\textbf{e,} Plot of the width-independent subparameters for $n_{\mathrm{eff, 0}}$, $\partial n_{\mathrm{eff}}/\partial\lambda_0$,and $\partial^2 n_{\mathrm{eff}}/\partial\lambda^2_0$ vs $V$.} 
    \label{fig:nominal_model_creation}
\end{figure*}
\color{black}\subsection{Sidewall Angle}
\color{black}We now consider how the parameter extraction behaves when used for waveguides with some sidewall angle. Up to this point, our simulations assumed the waveguides to have no sidewall angle.  Real waveguides, however, typically deviate from this ideal \cite{bogaerts2012silicon}. To study how our bias correction behaves under these conditions, a SOI waveguide with the same nominal (480~x~220~nm) design as before was simulated with a series of sidewall angles from 85 to 90 degrees as this is a range typical of foundries \cite{ye2005birefringence,xing2018accurate}. As only the aggregate behavior is being studied, width and thickness variations were not included. As seen in Fig. \ref{fig:sidewall_correction}a, the minimum of the error function optimized in the etch bias estimation step remain roughly constant for all considered sidewall angles. This results in very accurate predictions of the effective index from our model, even though the fundamental geometry is different. We interpret this as our optimization routine is picking an `equivalent' waveguide width that matches the extracted \color{black}$n_g$ \color{black} profile. This equivalent width always seems to result in a waveguide design with a similar confinement factor and effective index---and therefore behavior---as seen in Fig. \ref{fig:sidewall_correction}b.

\color{black}\subsection{Material Variation}\label{sec:material}
This method for increasing the accuracy of the guessed $n_{\mathrm{eff}}$ relies on the assumption that the material properties of the fabricated waveguides generally match the assumed material properties used in the simulation data used to construct the model. In practice, however, there can be a great deal of deviation between the assumed and actual optical properties of the waveguide materials. As a workaround, the authors suggest extracting and building a model based around the dispersion of the waveguide $\partial^2 n_{\mathrm{eff}}/\partial\lambda^2$, as this waveguide parameter can be extracted exactly from measurements. The nominal model of $\partial^2 n_{\mathrm{eff}}/\partial\lambda^2$ can then replace $n_g$ in \eqref{eq:error_function} to estimate the width of the measured device. This width can then be used in conjunction with simulation data to assign it an $n_{\mathrm{eff}}$ guess. Though the limits of such a technique are unclear to the authors, experimental results in Section \ref{sec:experiment} demonstrate to be effective enough for describing the $n_{\mathrm{eff}}$, loss, and thermo-optic effect for all measured device performance.

\color{black}\section{Extracting Local Parameter Variations}\label{sec:process_aware}

\begin{figure*}
    \centering
    \includegraphics{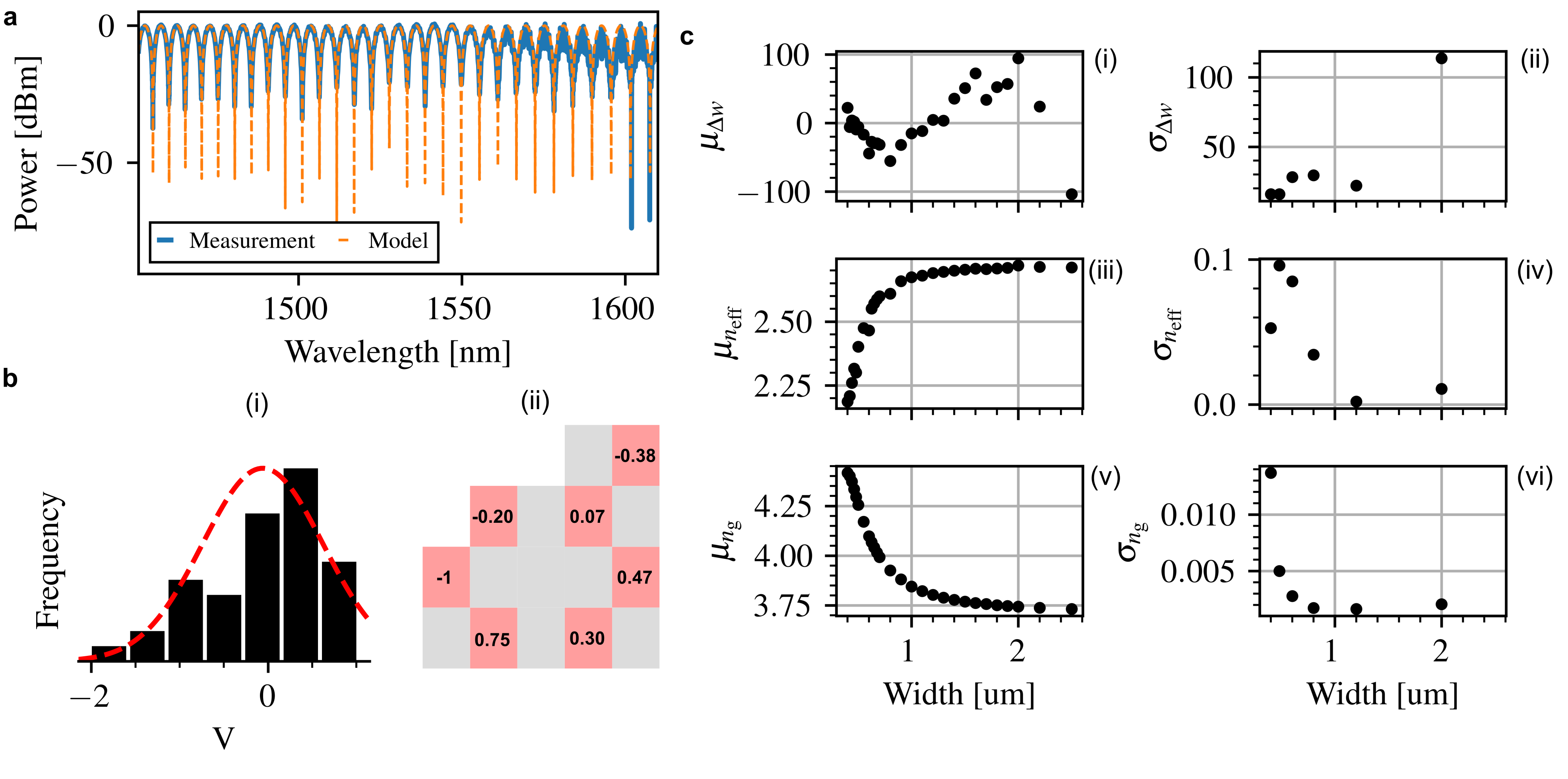}
    \color{black}\caption{\textbf{a}, Measured vs modeled optical spectrum of a 480~nm waveguide MZI with 
    $\Delta L = 100~\mu$m. \textbf{b}, (i) Histogram of extracted $V$ data along with its associated Gaussian distribution (red, dashed) overlaid on top. (ii) Spatial map of the average value for $V$ per measured die across the wafer. \textbf{c}, Mean and standard deviation of $\Delta w$, $n_{\mathrm{eff}}$, and $n_g$.}
    \label{fig:statistics}
\end{figure*}
Process variations (e.g. thickness variation, cladding and core index variations) will appear in our model as variations in the fifteen model parameters that comprise Eq. \eqref{eq:effective_index_model}. Capturing these variations requires the ability to extract their value locally, which cannot be done just by looking at the performance of any individual device. It is commonly assumed in prior literature that most process parameters slowly vary across the entire wafer \cite{xing2018accurate}. This assumption implies that the values of the parameters comprising our model also vary slowly across the wafer. The authors therefore propose analyzing the performance of several waveguide width designs in close proximity to each other to locally extract all of the fifteen model parameters. Each local extraction serves as the observations of each model parameter that are tracked across the entire wafer. 

The simplest way to create a statistical model is to treat each of fifteen sub-parameters as independent statistical variables. This is not ideal, however, as each additional variable drastically increases the number of required iterations for accurate Monte Carlo simulations. To minimize model complexity, we would like to represent each sub-parameter as a linear function of an ensemble of variables:
\begin{equation}
    p_{ni} = p_{ni,\mathrm{avg.}} + \vec{s}\cdot\vec{V}.
\end{equation}
$\vec{V}$ is the vector of variables that represent the process variations. Minimizing model complexity would be the equivalent of minimizing the size of $\vec{V}$. $\vec{s}$ describes the corresponding sensitivities of a given parameter to each element in $\vec{V}$. To minimize the size of $\vec{V}$, we leverage the fact that each extracted model parameter will be strongly correlated to one another. This is because the variations in each model parameter share common origins such as wafer thickness, annealing time, etc. We therefore propose using principal component analysis (PCA), a technique for transforming a number of possibly correlated variables into a smaller number of uncorrelated variables (i.e principal components) \cite{abdi2010principal}, to minimize model complexity. The chosen principal components are then the variables that make up $\vec{V}$. The chosen principal components are then the variables that make up $\vec{V}$. The number of components in $\vec{V}$ is flexible (see Appendix \ref{app:pca} for details). Since our waveguide geometry is primarily a function of two process variables---waveguide width and thickness---we use only the first principal component to preserve its physical interpretation. The result is a model of effective index as a function of width and our process variations--$\Delta w$, representing width variations and an additional variable we will call $V$, representing an aggregate of other process variations, including thickness variation:
\begin{equation}
    n_{\mathrm{eff, model}}(w+\Delta w, V).
\end{equation}
This full model of $n_{\mathrm{eff}}$ is then used in the local optimization and re-extraction of each measured device. The cost function is defined as the sum of the relative $n_{\mathrm{eff}}$ and $n_g$ errors to match both the measured fringe locations and FSR respectively.

Thus, we can employ a two-stage direct statistical compact model extraction procedure \cite{moezi2012statistical}. In the first stage, we use group extraction to obtain the complete set of fifteen parameters for a uniform device. In a second step, a subset of model parameters are re-extracted for each member of a large ensemble of devices measurements. This approach will be the most accurate representation of how device performance varies across the wafer without any presumption of variation source, statistical distribution, correlation, and the resulting model sensitivity to the variation. An inherent strength of this approach over others is that it is potentially useful for modeling other waveguide geometries as well. This potential is due to the model designer having the option of picking the number of principal components based on a physical assumption on the key process variables or optimize the percentage of explained parameter variance (see Appendix \ref{app:pca}). While further investigation would be required to confirm this, the methodology's flexibility holds a great deal of promise. 



\section{Experimental Demonstration}\label{sec:experiment}
We measured 7 reticles, each with 135 MZIs consisting of 27 different waveguide widths (w) from 400 nm to 2500 nm and 5 different arm length delays ($\Delta$L) from 100 nm to 500 nm, fabricated on a custom 300 mm full wafer through AIM Photonics (Fig. \ref{fig:nominal_model_creation}a). All 135 MZI were measured on reticle 2 while a smaller subset of 30 MZIs were measured on each of the remaining reticles, totaling 315 measured devices. Devices with the same waveguide width are placed adjacently to minimize the impact of local process variations on device performance. All MZIs have a nominal waveguide height of 220 nm, and grating couplers designed for quasi-TE polarization are utilized for optical I/O. The two arms of each MZIs consist of symmetric waveguide bends to mitigate the impact of bending on the $n_g$. For devices with waveguides beyond the single-mode cutoff width, Euler bends are used to maintain single mode operation and high mode isolation \cite{Rizzo:23}. A tunable laser was swept from 1450–1610 nm at a 10 pm resolution to characterize the transmission spectrum of each MZI.
\subsection{Nominal Extraction}
A nominal model $n_{\mathrm{eff}}$ is created by averaging the extracted parameters for all measured devices as shown in Fig. \ref{fig:nominal_model_creation}b. We apply the extraction method described in Section \ref{sec:extraction} to every collected transmission spectrum. A preliminary model is built using the simulation data described in Section \ref{sec:model} to estimate the expected device FSR for each waveguide width variation. This estimated FSR is then fed into a peak finding algorithm to extract the $n_g$ parameters, and then estimate fringe orders---and, therefore the $n_{\text{eff}}$---of each measured device. As the measured $n_g$ deviated a great deal from the simulated values, the technique described in Section \ref{sec:material} is employed where a preliminary model based on $\partial^2 n_{\mathrm{eff}}/\partial\lambda^2$ is created and used to estimate waveguide geometry to estimate the fringe order. All three Taylor-expansion parameters are then derived using \eqref{eq:B_derivation}-\eqref{eq:C_def}, and then averaged across for each width variation across the entire wafer to create a nominal experimental model. The extraction is then repeated locally for devices that are close in proximity to one another to extract local values for the model's sub-parameters (Fig. \ref{fig:nominal_model_creation}c). 

Fig. \ref{fig:nominal_model_creation}d shows that using \eqref{eq:variance_explained} this first principal component can explain 62.7\% of all variance in the sub-parameter values across the wafer. The authors determined that due to the clear connection between the $V$ and the three model parameters that determine device behavior as $w\rightarrow\infty$, this principal component was likely capturing width-independent sources of variance such as thickness variations (Fig. \ref{fig:nominal_model_creation}e). The authors will now show that this provides a model robust enough for capturing statistical behavior while preserving the goal for clear physical interpretation. 

\subsection{Statistical Extraction}\label{sec:stat_model_extraction}
\begin{figure}
    \centering
    \includegraphics{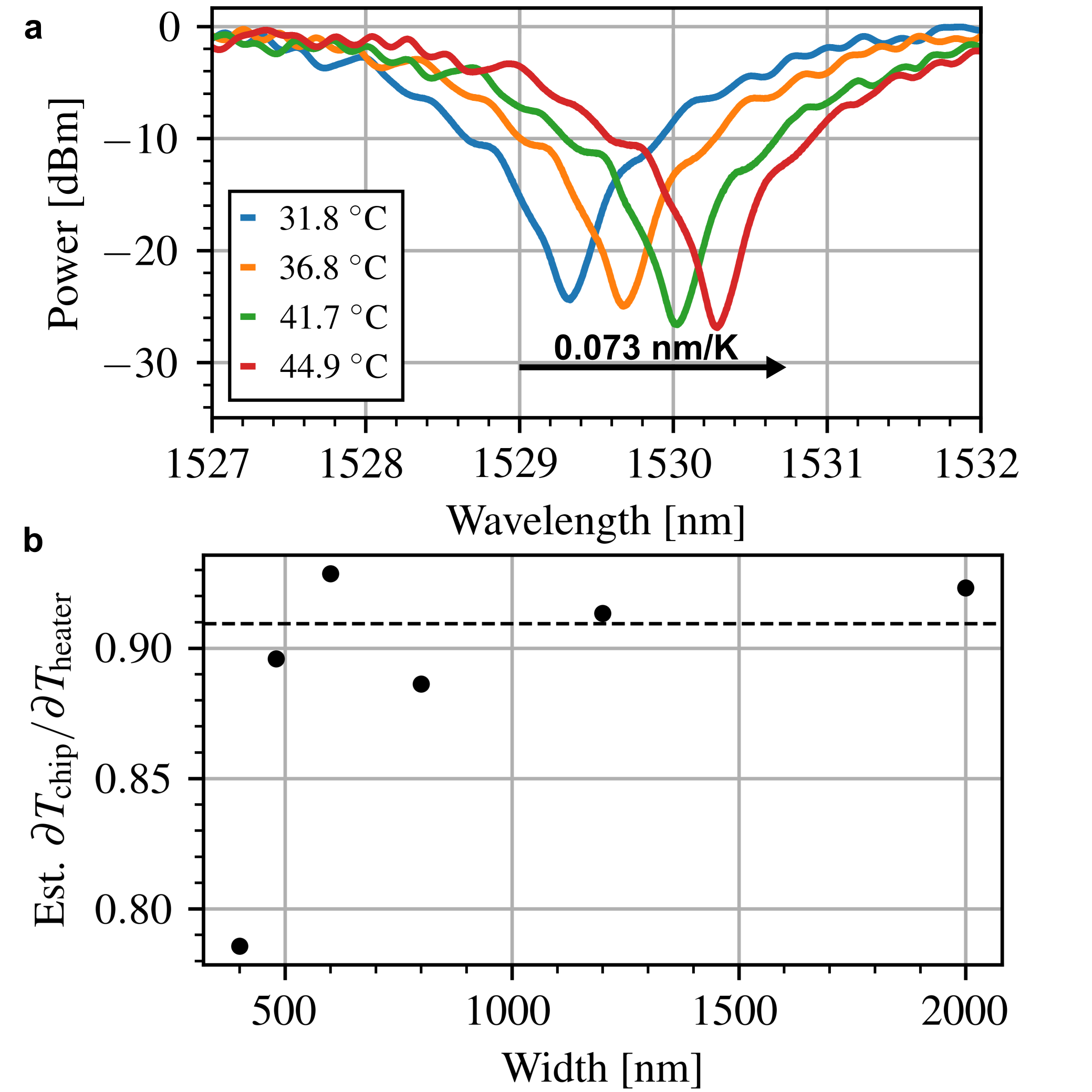}
    \color{black}\caption{\textbf{a}, Measured thermo-optic response of a measured MZI device. \textbf{b}, Extracted value of $\partial T_{\mathrm{chip}}/\partial T_{\mathrm{heater}}$ vs waveguide width using \eqref{eq:thermooptic_effect}.}
    \label{fig:heater_response}
\end{figure}
The sum of the relative $n_{\mathrm{eff}}$ and $n_g$ errors is optimized using the Nelder-Mead algorithm \cite{singer2009nelder}. The drawn waveguide width and the extracted $V$ from the local extraction performed in Fig. \ref{fig:nominal_model_creation}c are used as the initial guesses for $w$ and $V$. Despite the limited sample size of collected data, we can already see several notable preliminary statistical trends in $\Delta w$ as a result of our model. The model of $n_{\mathrm{eff}}$ extracted from each local optimization was found to result in a close agreement between the measured and modeled MZI performance. The extracted values for $V$ exhibits the intended physical behavior of process parameter that varies slowly across the wafer (Fig. \ref{fig:statistics}b). Local optimization yielded a total, average intra-die, and average local device standard deviation $\sigma_V$ of 0.603, 0.386, and 0.167 respectively, showing a correlation between device proximity and their extracted $V$ values. The mean values of $V$ for die both (i) in close proximity to each other and (ii) equidistant from the center of the wafer tend to be similar in value, as shown in the inset of Fig. \ref{fig:statistics}b. 

Decoupling the process variations of $V$ from the width variations $\Delta w$ enables extraction of width-dependent systemic effects, as shown in Fig. \ref{fig:statistics}c(i). Our method estimates that waveguide widths with smaller mean errors also tend to have smaller $\sigma_{\Delta w}$ (Fig. \ref{fig:statistics}c(ii)). This carries over as an explanation for why for $w=2 \mu$m, $n_{\mathrm{eff}}$ varies more than for $w=1.2 \mu$m, allowing insight on what waveguide geometry best minimizes both $\sigma_{n_{\mathrm{eff}}}$ and $\sigma_{n_{\mathrm{g}}}$. This sort of process insight for circuit designers is only possible due to the group benchmarking of all device performance within a localized area.

\subsection{Thermo-optic Effect Model Validation}\label{sec:thermo_validation}
To validate the thermo-optic effect model developed in Section \ref{sec:thermooptic}, we re-characterized the MZI transmission spectra from a single die of the chip shown in Fig. \ref{fig:nominal_model_creation}a. The thermal characterization was performed by adhering a Thorlabs TLK-H polyimide heater to the side of the chip stage. The heater was controlled by a Thorlabs TC200 Temperature Controller to set the heater temperature. Thermal paste was applied between the chip and the chip stage to minimize thermal resistance between the chip and the heater. The thermal response of one of the tested MZI is shown in Fig. \ref{fig:heater_response}. The fringe closest to 1550~nm is tracked at each temperature step and plotted against temperature to extract $\partial\lambda/\partial T$. This value is then compared to our predicted value for $\partial\lambda/\partial T$ gained by taking the derivative of $\lambda$ in \eqref{eq:resonance_condition} with respect to temperature
\begin{figure}
    \centering
    \includegraphics{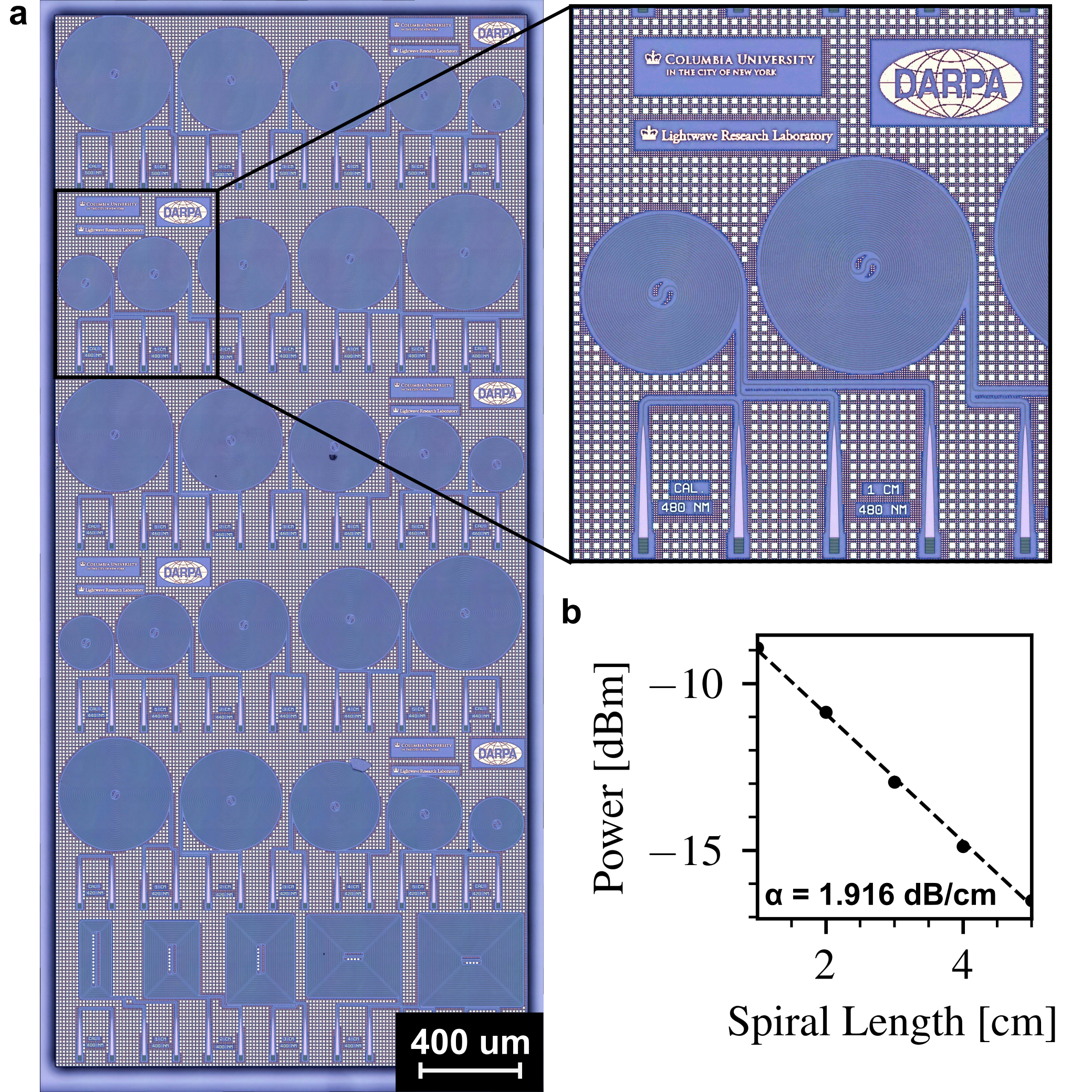}
    \color{black}\caption{\textbf{a}, Microscopic image of a die with waveguide spiral test structures for measuring width-dependent loss. Inset: magnified image showing three of the test structures. \textbf{b}, Propagation loss measurement and fit data for a 440~nm waveguide.}
    \label{fig:loss_chip}
\end{figure}
\begin{equation}\label{eq:fringe_shift}
    \frac{\partial\lambda}{\partial T_{\mathrm{chip}}}=\frac{\displaystyle\frac{\Delta L}{m}\frac{\partial n_{\mathrm{eff}}}{\partial T_{\mathrm{chip}}} \frac{\partial T_{\mathrm{chip}}}{T_{\mathrm{heater}}}}{\displaystyle 1 - \frac{\Delta L}{m}\frac{\partial n_{\mathrm{eff, model}}}{\partial \lambda}},
\end{equation}
where $m$ is the order of the tracked fringe, $\Delta L$ is the path length difference between the two arms, and $\partial T_{\mathrm{chip}}/\partial T_{\mathrm{heater}}$ represents the heat transfer efficiency from the heater to the chip itself. This last term is included as the authors only know the temperature of the resistive heater rather than the chip temperature itself. We know $\partial T_{\mathrm{chip}}/\partial T_{\mathrm{heater}}\leq1$ as heater cannot raise the temperature of the chip to a value higher than its own. The extracted parameters for $\Delta w$ and $V$ are used in calculating \eqref{eq:fringe_shift}. 

On average, the measured thermo-optic effect was found to be 0.91$\times$ our model's~\eqref{eq:thermooptic_effect} prediction. This value was found to be independent of waveguide geometry with the exception of 400~nm (Fig. \ref{fig:heater_response}b). This error for 400~nm is assumed to be because this width is close enough to the cutoff condition for our model to lose accuracy. In contrast, the measured thermo-optic effect was 1.21$\times$ the previously reported model's prediction. This implies that either the chip's change in temperature is greater than the heater's or the previously reported model is incorrect. The change in temperature of the PIC can only ever be smaller than the heater's temperature delta, making the old model's prediction clearly nonphysical.
\subsection{Scattering Loss Model Validation}
\begin{figure}
    \centering
    \includegraphics{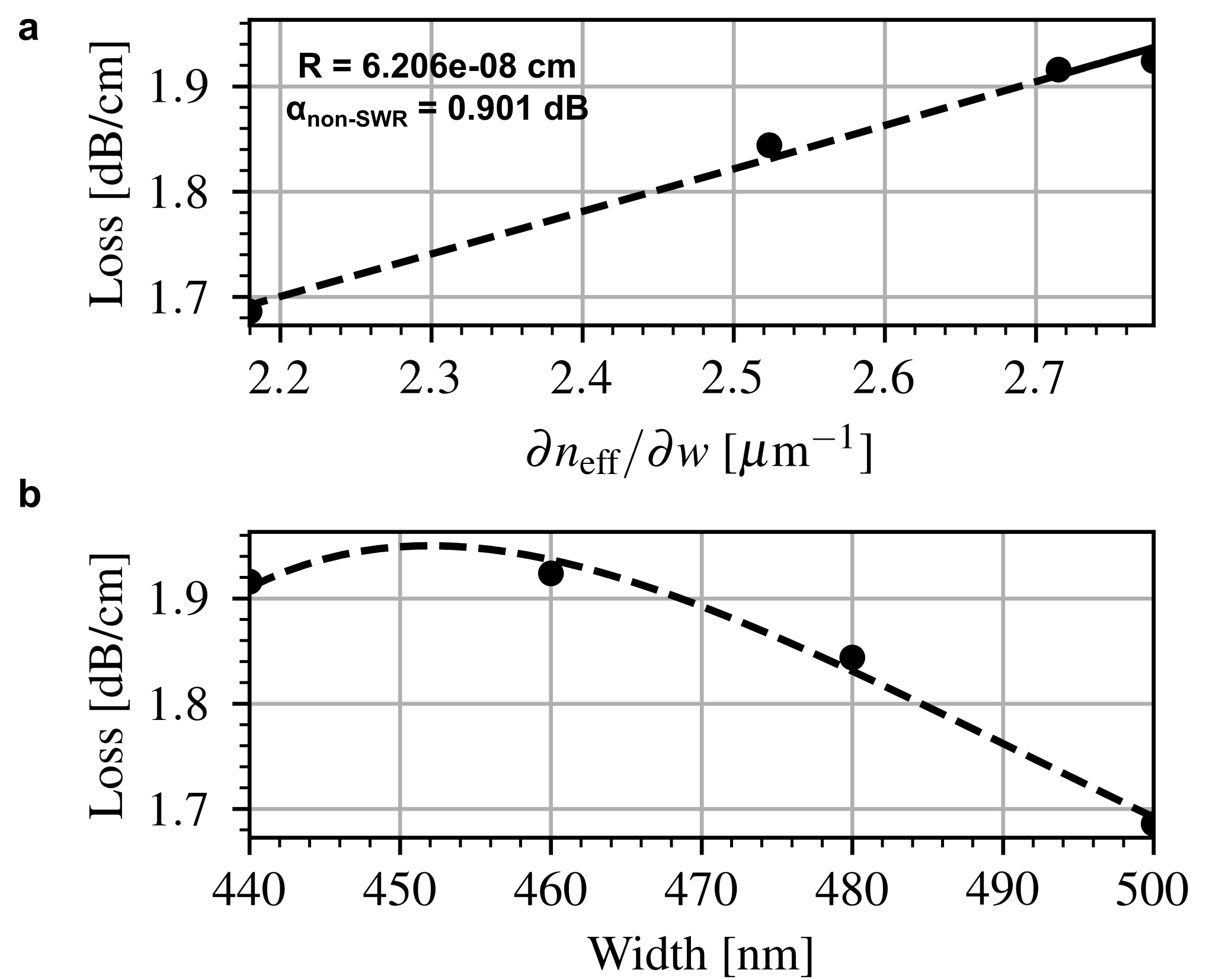}
    \color{black}\caption{\textbf{a,} Plot of measured (scatter) and modeled (line) propagation loss vs $\partial n_{\mathrm{eff}}/\partial w$. Slope of fit represents $R$ in \eqref{eq:swr_loss_fit}, while the intercept represents non-SWR loss. \textbf{b,} Plot of measured (scatter) and modeled (line) propagation loss vs waveguide width.}
    \label{fig:loss_fit}
\end{figure}
To validate the scattering loss model, we measured a die with 25 spiral loss structures consisting of 5 different waveguide widths (w) from 400~nm to 500~nm and 5 different spiral lengths ($\Delta$L) from 1 cm to 5 cm, fabricated on a custom 300 mm full wafer through AIM Photonics (Fig. \ref{fig:loss_chip}). Again, all spiral structures have a nominal waveguide height of 220 nm, and grating couplers designed for quasi-TE polarization are utilized for optical I/O. The losses of each spiral length were recorded, and then fit to a linear equation. The slope of the this fit was taken to be the propagation loss associated with each waveguide width. The results of our model fit are shown in Fig. \ref{fig:loss_fit}. The model built in Section \ref{sec:stat_model_extraction} was used to build a model of $\partial n_{\mathrm{eff}}/\partial w$. Fitting our modeled $\partial n_{\mathrm{eff}}/\partial w$ to the measured propagation loss yields proportionality constant of $R=\num{6.206e-8}$~cm and (Fig. \ref{fig:loss_fit}a). The intercept of the loss fit is interpreted as the aggregate non-SWR loss, with a value of 0.901~dB. Fig. \ref{fig:loss_fit}b shows the excellent agreement between our model and the data, predicting the similar propagation losses of both the 440~nm and 460~nm waveguides. As mentioned in Section \ref{sec:stat_model_extraction}, both 400 and 420~nm waveguide widths are likely near the cutoff condition. Since \eqref{eq:swr_loss_fit} is only valid sufficiently far away from this condition, those data points are not included in the plot.
\color{black}\subsection{Verilog-A Implementation}
\begin{figure}
    \centering
    \includegraphics{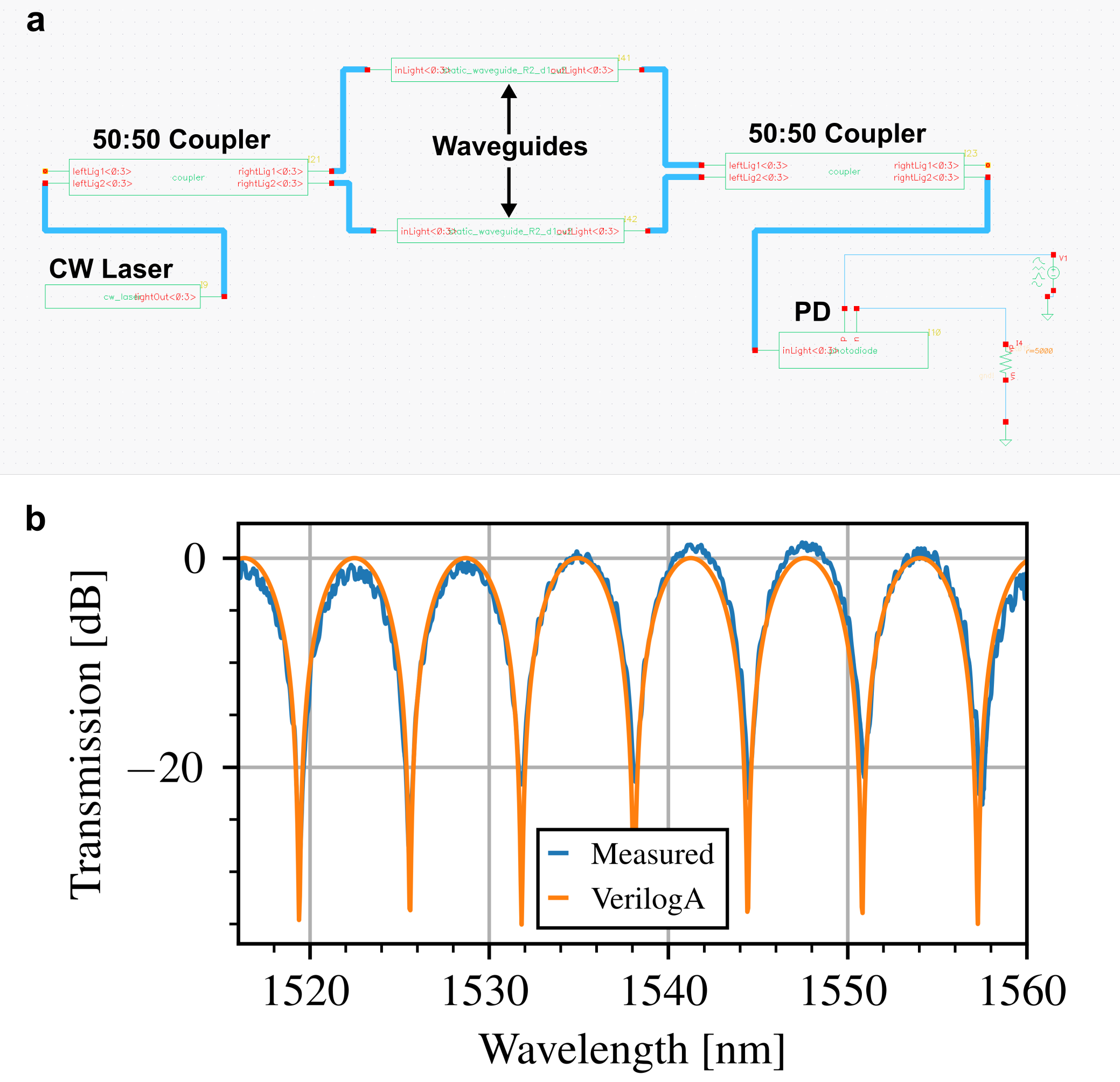}
    \caption{\textbf{a}, Cadence Virtuoso schematic of the MZI test circuit. All circuit models were written in Verilog-A. The optical stimulus is provided by a continuous-wave (CW) Laser and detected with a photodetector (PD). \color{black}\textbf{b}, Comparison of measured and simulated performance for an MZI with $w=2~\mu$m and $\Delta L=100~\mu$m.}
    \label{fig:verilogA}
\end{figure}

To demonstrate its compatibility with electronic-photonic co-simulation, the circuit model was implemented in Verilog-A within Cadence Virtuoso (Fig. \ref{fig:verilogA}a). As Verilog-A does not inherently support optical signals, some compatibility code as well as a small library of photonic device models were built based upon on previously reported demonstrations \cite{kononov2013modeling,sorace2015electro,leu2018integrated}.

  
\color{black}\section{Conclusion}
In summary, we have demonstrated a novel compact model that can greatly expand the accuracy of circuit-level simulation capabilities of silicon PICs. In contrast to prior work that focused on providing metrology information that could be use to fabrication engineers \cite{Xing:20, lu2017performance, boning2022variation}, we present this PDK model as a tool suitable for true-to-measurement circuit simulation and optimization. By leveraging this underlying physical behavior and locally extracting process variations by performing group extraction, we have demonstrated a framework for building a model of $n_{\mathrm{eff}}$ that is entirely driven by measurement data. This model was shown to accurately describing the phase, loss, and thermo-optic behavior of the measured integrated waveguides over 4$\times$ the optical bandwidth and over 80$\times$ the range of waveguides widths reported in prior work. We envision that the advancement over prior demonstrations this work represents can support the development of waveguide-based PDK components and enable the robust optimization of next generation PICs. 

\color{black}
\section{Acknowledgements}
This work was supported in part by the U.S. Advanced Research Projects Agency--Energy under ENLITENED Grant DE-AR000843 and in part by the U.S. Defense Advanced Research Projects Agency under PIPES Grant HR00111920014. The authors thank AIM Photonics for chip fabrication. 

\appendix
\subsection{Derivation of Thermo-Optic Model}\label{app:thermal}

Starting from \eqref{eq:wave_equation_perturbed} from \cite{yariv2007photonics}, the effect of a thermal perturbation on the effective index is investigated. Carrying out this perturbation and following the chain rule yields:
\begin{equation}
    2\beta\frac{\partial\beta}{\partial T}=2\Gamma_{\text{core}}\frac{\omega^2}{c^2}n_{\text{core}}\frac{\partial n_{\text{core}}}{\partial T} + 2\Gamma_{\text{clad}}\frac{\omega^2}{c^2}n_{\text{clad}}\frac{\partial n_{\text{clad}}}{\partial T}.
\end{equation}
Noting that $\beta=\omega n_{\text{eff}}/c$ and inserting above, the relationship simplifies to \eqref{eq:final_thermal_relationship}. Combining with \eqref{eq:effective_index_model} yields:
\begin{subequations}\label{eq:final_thermal_relationship}
    \begin{align}
        n_{\text{eff}} = n_{\text{eff, T$_0$}}+\frac{\partial n_{\text{eff}}}{\partial T}(T-T_0)\\
        \frac{\partial n_{\text{eff}}}{\partial T} = \Gamma_{\text{core}}\frac{n_{\text{core}}}{n_{\text{eff}}}\frac{\partial n_{\text{core}}}{\partial T}+\Gamma_{\text{clad}}\frac{n_{\text{clad}}}{n_{\text{eff}}}\frac{\partial n_{\text{clad}}}{\partial T}.
    \end{align}
\end{subequations}
It is noted that $n_{\text{eff}}$ appears on both sides of the equation. Multiplying both sides by effective index yields a quadratic equation whose solution is:
\begin{subequations}\label{eq:full_thermooptic_effect}
    \begin{align}
        n_{\text{eff}}=\frac{n_{\text{eff, T$_0$}}}{2}+\frac{1}{2}\sqrt{n_{\text{eff, T$_0$}}^2+4n^{'} (T-T_0)}\\
        n^{'} = \Gamma_{\text{core}}n_{\text{core}}\frac{\partial n_{\text{core}}}{\partial T}+\Gamma_{\text{clad}}n_{\text{clad}}\frac{\partial n_{\text{clad}}}{\partial T}.
    \end{align}
\end{subequations}
The expression can be simplified by noting that $n_{\text{eff, T$_0$}}^2\gg4n^{'}$ for typical values for the thermo-optic coefficients. Understanding this, it is clear that the behavior of the square root term is approximately linear. The 1st order Taylor expansion of the square root term is:
\begin{equation}\label{eq:taylor_thermal}
    n_{\text{eff, T$_0$}}+\frac{1}{2}\frac{4n^{'}}{\sqrt{n_{\text{eff, T$_0$}}^2+4n^{'} (T-T_0)}}(T-T_0).
\end{equation}
Noting again that $n_{\text{eff, T$_0$}}^2\gg4n^{'}$, \eqref{eq:taylor_thermal} simplifies to:
\begin{equation}
    n_{\text{eff, T$_0$}}+\frac{2n^{'}}{n_{\text{eff, T$_0$}}}(T-T_0).
\end{equation}
Replacing the square root term in \eqref{eq:full_thermooptic_effect} with this expression and simplifying will then yield \eqref{eq:thermooptic_effect}.
\color{black}\subsection{Principal Component Analysis}\label{app:pca}
To start, we form a matrix $\textbf{X}$ our of our list of local sub-parameter extractions, where each column represents a model parameter and each row is an observation of said parameter: 
\begin{equation}
    \textbf{X} =
  \left[ {\begin{array}{cccc}
    \frac{\partial^0n_{\mathrm{eff}}}{\partial\lambda^0}_{0,1} & \frac{\partial^0n_{\mathrm{eff}}}{\partial\lambda^0}_{1,1} & \cdots & \frac{\partial^2n_{\mathrm{eff}}}{\partial\lambda^2}_{4,1}\\
    \frac{\partial^0n_{\mathrm{eff}}}{\partial\lambda^0}_{0,2} & \frac{\partial^0n_{\mathrm{eff}}}{\partial\lambda^0}_{1,2} & \cdots & \frac{\partial^2n_{\mathrm{eff}}}{\partial\lambda^2}_{4,2}\\
    \vdots & \vdots & \ddots & \vdots\\
    \frac{\partial^0n_{\mathrm{eff}}}{\partial\lambda^0}_{0,n} & \frac{\partial^0n_{\mathrm{eff}}}{\partial\lambda^0}_{1,n} & \cdots & \frac{\partial^2n_{\mathrm{eff}}}{\partial\lambda^2}_{4,n}\\
  \end{array} } \right].
\end{equation}
A covariance matrix \textbf{S} is then created from $\textbf{X}$ and find its eigenvectors:
\begin{equation}
    \textbf{S} =
\left[ {\begin{array}{c}
		\vec{v}_0 \\
		\vec{v}_1 \\
    		\vdots\\
   		\vec{v}_n\\
  \end{array} } \right]
  \left[ {\begin{array}{cccc}
    \lambda_0 & 0 & \cdots & 0\\
    0 & \lambda_1 & \cdots & 0\\
    \vdots & \vdots & \ddots & \vdots\\
    0 & 0 & \cdots & \lambda_n\\
  \end{array} } \right]
\left[ {\begin{array}{c}
		\vec{v}_0 \\
		\vec{v}_1 \\
    		\vdots\\
   		\vec{v}_n\\
  \end{array} } \right]^{-1},
\end{equation}
where $\left[v_0,v_1,\cdots,v_n\right]$ lists the eigenvectors and $\left[\lambda_0,\lambda_1,\cdots,\lambda_n\right]$ are their associated eigenvalues. The eigenvectors of the correlation matrix represent the directions of the axes where there is the most variance (i.e. the most information). Each eigenvalue $\lambda_i$ is proportional to how much variance is captured by its associated principal component $v_i$. Picking the eigenvectors with the largest eigenvalues allows us to reduce data dimensionality at the expense of some accuracy. The percentage of variability explained by a principal component is calculated as
\begin{equation}\label{eq:variance_explained}
    \frac{\sum_{i=0}^M\lambda_{i}}{\sum_{i=0}^N\lambda_{i}},
\end{equation}
where $\lambda_i$ is the eigenvalue for each eigenvector, $M$ is the number of principal components the designer has chosen to include, and $N$ is the maximum number of principal components.\color{black}

\bibliographystyle{IEEEtran}
\bibliography{bibliography}

\begin{IEEEbiographynophoto}{Aneek James}
received his B.S. in Electrical and Electronics Engineering from the University of Georgia, Athens, GA in 2017 and his M.S., and M.Phil., in Electrical Engineering from Columbia University, New York, NY in 2019 and 2021, respectively. He is working as a Ph.D. candidate in Electrical Engineering in the Lightwave Research Laboratory under Professor Keren Bergman. His research interests include modeling fabrication variations in silicon photonic devices, as well as the testing and automated control of silicon photonic systems for high-throughput optical interconnects.
\end{IEEEbiographynophoto}

\begin{IEEEbiographynophoto}{Anthony Rizzo}
received his B.S. in Physics from Haverford College, Haverford, PA in 2017 and his M.S., M.Phil., and Ph.D., all in Electrical Engineering, from Columbia University, New York, NY in 2019, 2021, and 2022, respectively. He completed his doctoral research in the Lightwave Research Laboratory at Columbia University under Professor Keren Bergman, where he led the first demonstration of data transmission using an integrated Kerr frequency comb source and silicon photonic transmitter. He is currently a Research Scientist at the Air Force Research Laboratory (AFRL) Information Directorate in Rome, NY, with a focus in large-scale silicon photonic systems for quantum information processing and artificial intelligence.
\end{IEEEbiographynophoto}

\begin{IEEEbiographynophoto}{Yuyang Wang}
received the B.Eng. degree in electronic engineering from Tsinghua University, Beijing, China in 2015, and the M.S. and PhD degrees in computer
engineering from the University of California, Santa Barbara (UCSB), CA, USA, in 2018 and 2021 respectively. He is currently a post-doctoral researcher in the Lightwave Research Laboratory under Professor Keren Bergman. He was a Design Engineering Intern at Cadence Design Systems in 2018 and a Visiting Intern at the Hong Kong University of Science and Technology in 2019. His research interests include variation-aware modeling, design, and optimization of silicon photonic interconnects and systems.
\end{IEEEbiographynophoto}

\begin{IEEEbiographynophoto}{Asher Novick}
received his M.Eng. and B.S. degrees in Electrical and Computer Engineering from Cornell University, Ithaca, NY, USA, in 2016 and 2015,respectively. Between 2016 and 2019, he was at Panduit’s Fiber Research Lab, where he researched and developed new patentable technologies for optical fiber-based communication in data center and enterprise applications. He is currently working toward his Ph.D. degree in Electrical Engineering in the Lightwave Research Laboratory at Columbia University in the City of New York. His current research interest is in the modeling, design, and testing of silicon photonic systems and devices for scalable and efficient link architectures.
\end{IEEEbiographynophoto}

\begin{IEEEbiographynophoto}{Songli Wang}
received his B.Eng. in Optoelectronic Information Science and Engineering from Harbin Institute of Technology, Harbin, China, in 2019 and his M.S. in Electrical Engineering from Columbia University, New York, NY in 2020. He is currently working towards the Ph.D. degree in Electrical Engineering in the Lightwave Research Laboratory at Columbia University. His current research interests include modeling, design and testing of silicon photonic devices and systems.
\end{IEEEbiographynophoto}

\begin{IEEEbiographynophoto}{Robert Parsons}
received the B.S. degree in biomedical engineering from George Washington University, Washington, D.C., USA, and the M.S. degree in electrical engineering from Columbia University, New York, NY, USA, in 2020 and 2022, respectively. He is currently working toward the Ph.D. degree in electrical engineering with the Lightwave Research Laboratory, Columbia University under Professor Keren Bergman. His research interests include the modeling, testing, and co-optimization of link architectures and constituent silicon photonic devices for high-bandwidth, energy-efficient optical interconnects.
\end{IEEEbiographynophoto}

\begin{IEEEbiographynophoto}{Kaylx Jang}
received his B.S. in Electrical Engineering from the University of California, Irvine, CA in 2020 and his M.S. in Electrical Engineering from Columbia University in 2022. In 2019 and 2020, he interned in the testing department at Ayar Labs and in 2021 did a co-op in the Silicon Photonics Design team at Nokia (former Elenion). He is working as a Ph.D. student in Electrical Engineering in the Lightwave Research Lab under Professor Keren Bergman.  His research interests include modeling, design, and testing of high-performance silicon photonic devices for energy efficient and scalable link architectures.
\end{IEEEbiographynophoto}

\begin{IEEEbiographynophoto}{Maarten Hattink}
is a graduate student with Columbia University, New York, NY, USA. He received the B.S. and M.S. degrees from the Eindhoven University of Technology, The Netherlands, in 2015 and 2017, respectively. While pursuing these degrees, he worked at Prodrive Technologies B.V. as a Software and FPGA Engineer. He is currently working toward the Ph.D. degree and his research interest lies in photonic device integration and thermal control.
\end{IEEEbiographynophoto} 

\begin{IEEEbiographynophoto}{Keren Bergman (S’87–M’93–SM’07–F’09)}
received the B.S. degree from Bucknell University, Lewisburg, PA, in 1988, and the M.S. and Ph.D. degrees from the Massachusetts Institute of Technology, Cambridge, in 1991 and 1994, respectively, all in electrical engineering. Dr. Bergman is currently a Charles Batchelor Professor at Columbia University, New York, NY, where she also directs the Lightwave Research Laboratory. She leads multiple research programs on optical interconnection networks for advanced computing systems, data centers, optical packet switched routers, and chip multiprocessor nanophotonic networks-on-chip. Dr. Bergman is a Fellow of the IEEE and Optica.
\end{IEEEbiographynophoto}

\end{document}